\documentclass[aps,prl,reprint,superscriptaddress,showpacs,showkeys]{revtex4-1}

\usepackage[]{hyperref} 
\usepackage{amsmath}
\usepackage{amsfonts}
\usepackage{textcomp}
\usepackage[]{graphicx}
\usepackage{epstopdf}
\usepackage{mathrsfs}
\usepackage{amsbsy}
\usepackage{cleveref}

\crefformat{equation}{Eq.~(#2#1#3)}
\crefformat{figure}{Fig.~#2#1#3}

\Crefformat{equation}{Equation~(#2#1#3)}
\Crefformat{figure}{Figure~#2#1#3}

\hypersetup{pdfauthor={Walasik, Silahli, Litchinitser},pdftitle={OAM beams in colloids}}

\begin{document}

\title{Trajectories and orbital angular momentum of necklace beams \\ in nonlinear colloidal suspensions}

\author{Wiktor Walasik}
\affiliation{Department of Electrical Engineering, University at Buffalo, The State University of New York, Buffalo, New York 14260, USA}
\author{Salih Z. Silahli}
\affiliation{Department of Electrical Engineering, University at Buffalo, The State University of New York, Buffalo, New York 14260, USA}
\author{Natalia M. Litchinitser}
\affiliation{Department of Electrical Engineering, University at Buffalo, The State University of New York, Buffalo, New York 14260, USA}
\email[]{wiktorwa@buffalo.edu}

\graphicspath{ {./Figures/} }

\date{\today}

\begin{abstract}
Recently, we have predicted that the modulation instability of optical vortex solitons propagating in nonlinear colloidal suspensions with exponential saturable nonlinearity leads to formation of necklace beams (NBs) [S.~Z.~Silahli, W.~Walasik and N.~M.~Litchinitser, Opt.~Lett., \textbf{40}, 5714 (2015)]. Here, we investigate the dynamics of NB formation and propagation, and show that the distance at which the NB is formed depends on the input power of the vortex beam. Moreover, we show that the NB trajectories are not necessarily tangent to the initial vortex ring, and that their velocities have components stemming both from the beam diffraction and from the beam orbital angular momentum. We also demonstrate the generation of twisted solitons and analyze the influence of losses on their propagation. Finally, we investigate the conservation of the orbital angular momentum in necklace and twisted beams. Our studies, performed in ideal lossless media and in realistic colloidal suspensions with losses, provide a detailed description of NB dynamics and may be useful in studies of light propagation in highly scattering colloids and biological samples.
\end{abstract}

%


\maketitle

Structured light, especially optical vortices carrying orbital angular momentum (OAM)~\cite{Allen92}, attracted a growing attention in last two decades. A particular interest was paid to generation of spiraling light patterns naturally stemming from the twisted nature of OAM beams. In the linear regime, spiraling beams can be created by interference of two vortices with opposite charges and different longitudinal propagation constants~\cite{CHAVEZCERDA1996,Schechner96,ABRAMOCHKIN1997,Tervo:01,Schulze15,Wei-Ping10}. In nonlocal nonlinear media, stable spiraling solitons were found. They can originate from an optical fields carrying a nonzero OAM~\cite{Tikhonenko:95,Tikhonenko96,Mihalache02,Buccoliero07,Buccoliero:08,Liang13,Wen16}, or a collision of skewed solitons that do not carry OAM~\cite{Poladian91,Shih97,Belic99,Buryak99,Lem-Carrillo14}. A broad review of vortex solitons and interaction of such beams can be found in Refs.~\cite{Desyatnikov05,Rotschild06}. Another type of beams that originate from optical vortices are necklace beams (NB)---beams with azimuthal modulation of intensity profile. Such beams can be created by interference of optical vortices with different charges~\cite{Soljacic98,Soljacic00,Soljacic01,Grow07}, or can be a result of modulation instability (MI) induced splitting of a vortex beam~\cite{Firth97,Skryabin98,Vinçotte06,Vuong06,Silahli:15} or a super-Gaussian beam~\cite{Grow:06}. Solitons building a NB have been shown to either escape from the NB ring following trajectories tangent to the NB ring~\cite{Firth97,Skryabin98} or fuse and create a vortex pattern~\cite{He:07}, depending on their arrangement and phase distribution.

Colloidal suspensions (CSs) constitute a useful platform to study nonlinear light-matter interactions. 
Many interesting effects in CSs have been reported, such as optical trapping of dielectric particles~\cite{Ashkin70,He1995,Simpson1996,Volke-Sepulveda2002,Gordon07} that found applications in optical tweezing, and creation of artificial nonlinear media~\cite{Ashkin:82,Smith:82}. The initial prediction of the Kerr-like nature of the nonlinearity of CSs was later refined. It was shown that the nonlinearity of CSs has exponential character and can be either supercritical, in case of particles with positive polarizability, or saturable, for negative polarizability particles~\cite{El-Ganainy:07b,Lee:09}. Propagation, stability, and interactions of Gaussian beams~\cite{El-Ganainy:07,Man13,Fardad:13} and optical vortices~\cite{Silahli:15} in nonlinear CSs have been studied extensively.

In our previous work, we have investigated the propagation of vortex beams in nonlinear CSs and have shown that NBs can be generated via the MI~\cite{Silahli:15}. The initial vortex beam corresponded to a stable solution of a nonlinear Schr\"odinger equation defined by a fixed vortex radius and associated power level. In this work, we demonstrate generation of NBs from vortices that are not stationary solutions that offer more freedom in choice of the input beam parameters. This freedom allows us to modify the vortex power, keeping the radius fixed and observe how it influences the distance at which the MI onsets. Moreover, we study the trajectories of the solitons forming the NB and analyze the evolution of the NB profile in the transverse plane, depending on the charge of the vortex and its radius, for fixed power.
Finally, we demonstrate generation of twisted solitons resulting from the fusion of two solitons generated by a vortex, and investigate the OAM conservation in these phenomena.

We study the propagation of beams carrying the OAM in a CS built of air bubbles with refractive index $n_p = 1$ uniformly distributed in water with the refractive index $n_b = 1.33$. We note that in the laboratory experiments, one can use polytetrafluoroethylene (PTFE) particles dispersed in glycerin--water solution to realize a stable negative polarizability CS. The volumetric filing fraction of this solution is $f = 1$\textperthousand\, and the radius of the air bubbles is taken as $r_p = 50$~nm. The suspension is assumed to be at the room temperature $T=293$~K and the incoming free-space wavelength of light is $\lambda_0 = 2\pi/k_0 = 532$~nm. The nonlinearity in such a system has exponentially saturable character~\cite{El-Ganainy:07b} and the propagation of the slowly varying envelope of the electric field $\phi$ can be described by 
\begin{equation}
i\frac{\partial \phi}{\partial z} + \frac{\nabla^2_{\perp}\phi}{2 k_0 n_b} + \left[k_0(n_p-n_b)V_p + \frac{i}{2}\sigma\right]\rho_0 e^{\frac{\alpha}{4k_BT}|\phi|^2}\phi =0, \nonumber
\end{equation}
where $z$ denotes the propagation distance, $\nabla^2_{\perp}$ is the transverse Laplacian, $V_p=4\pi r_p^3/3$ is the volume of the particle, $\sigma$ is the scattering cross section responsible for loss, $\rho_0 =f/V_p $ is the unperturbed particle concentration, $\alpha$ is the particle polarizability, and $k_B$ is the Boltzmann constant.

\begin{figure}[!t]
	\includegraphics[width = 0.49\columnwidth, clip = true, trim = {60 0 60 50}]{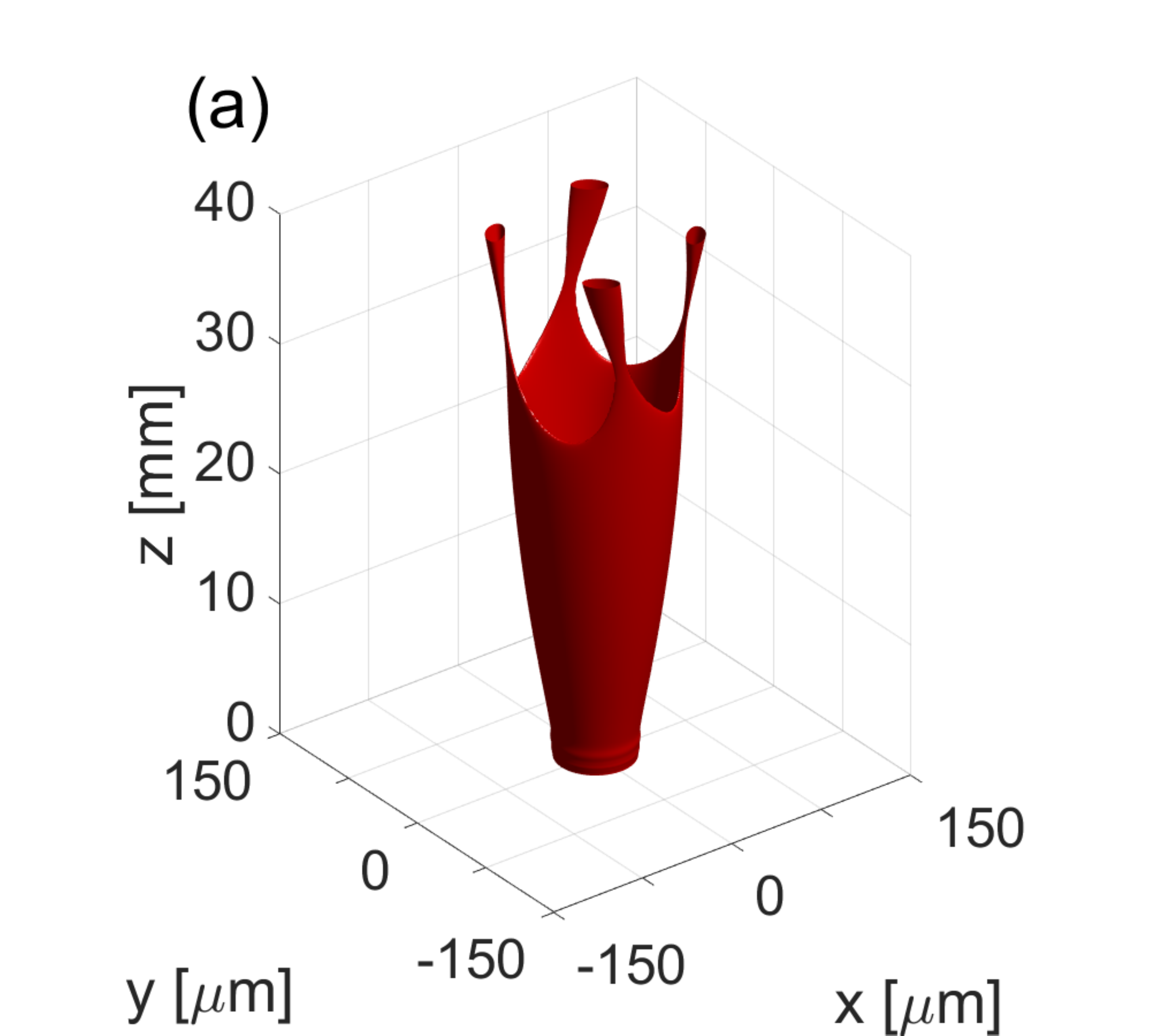}
	\includegraphics[width = 0.49\columnwidth, clip = true, trim = {60 0 60 50}]{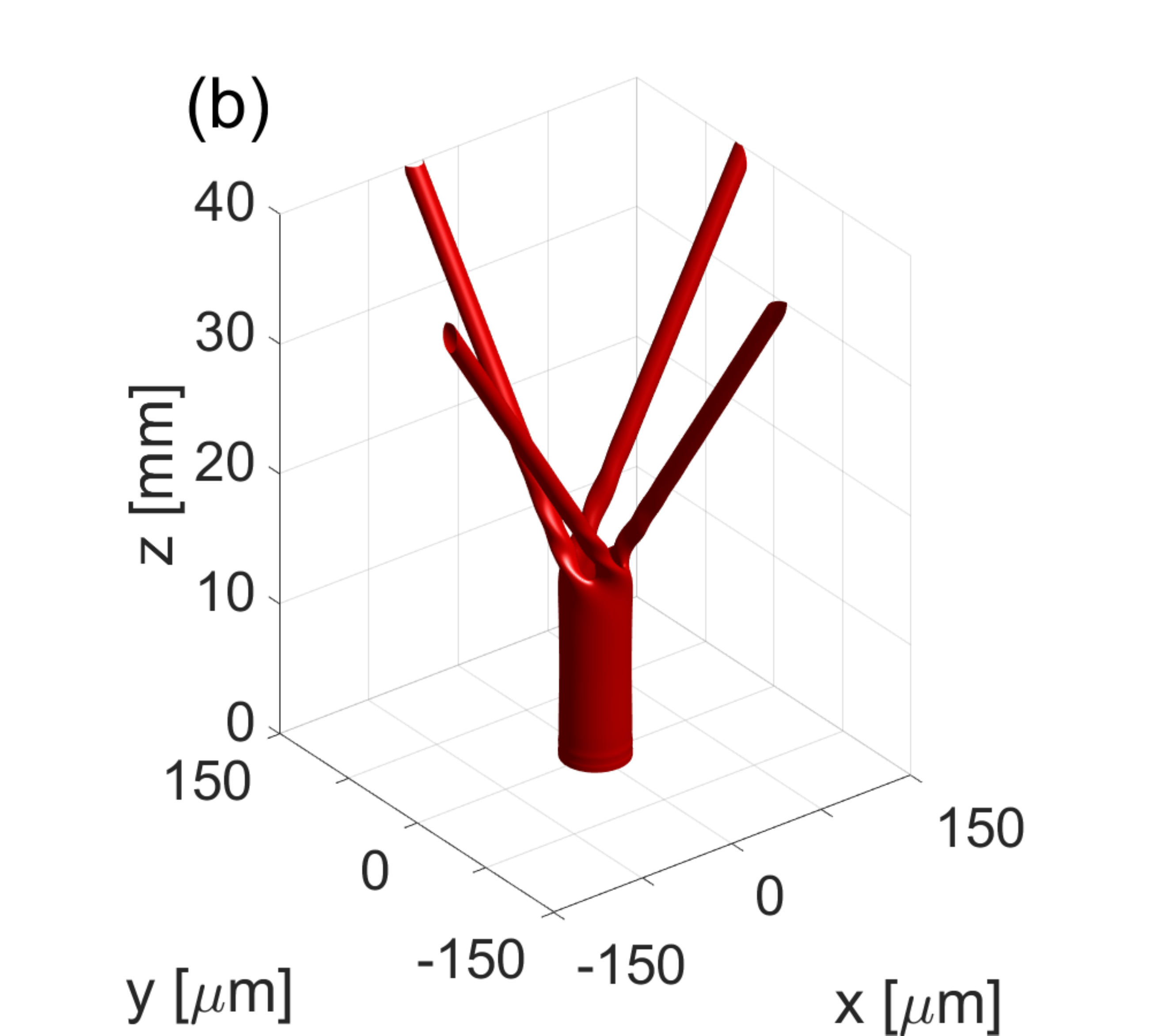}
	\includegraphics[width = 0.49\columnwidth, clip = true, trim = {60 0 60 50}]{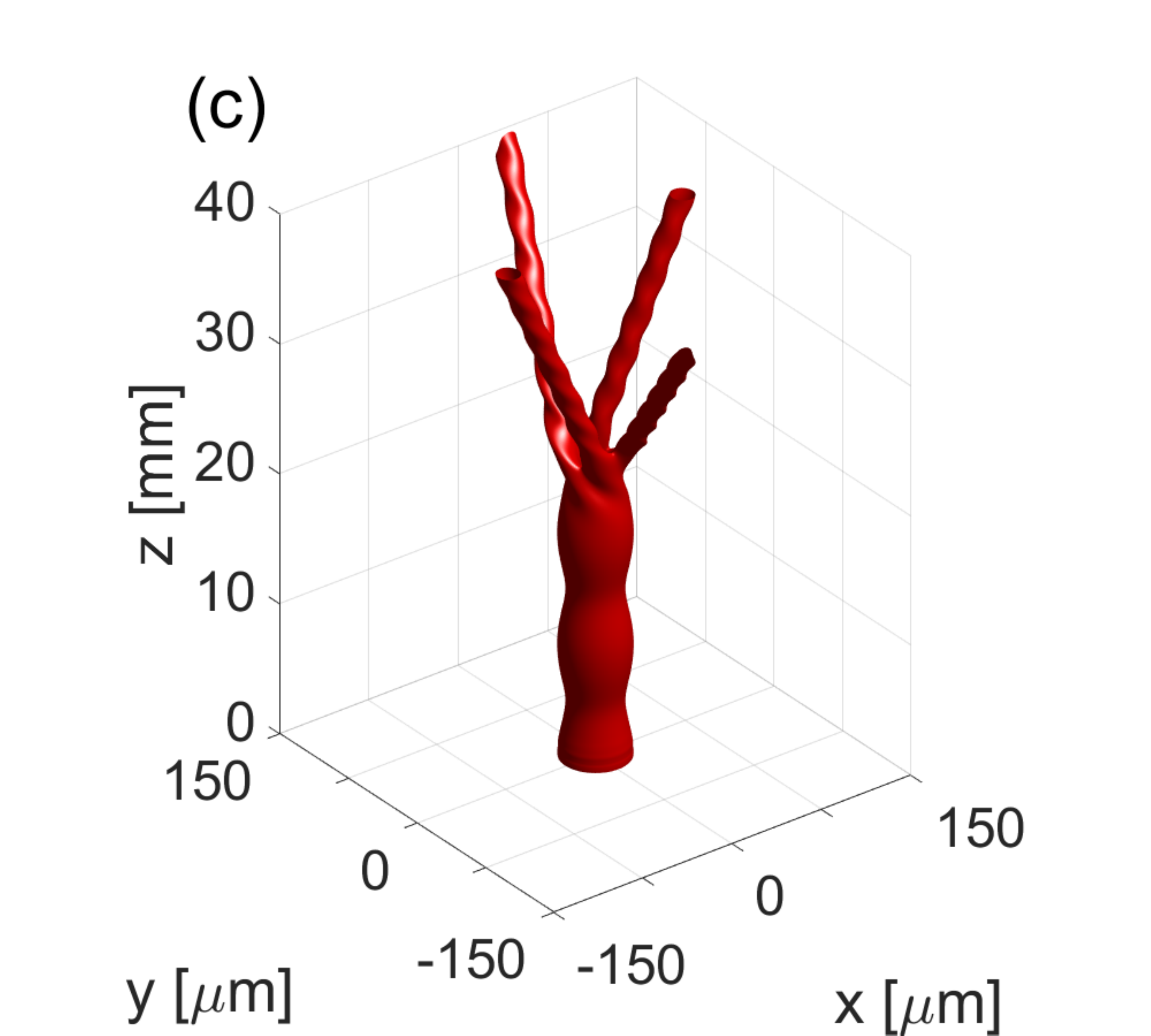}
	\includegraphics[width = 0.49\columnwidth, clip = true, trim = {0 0 150 0}]{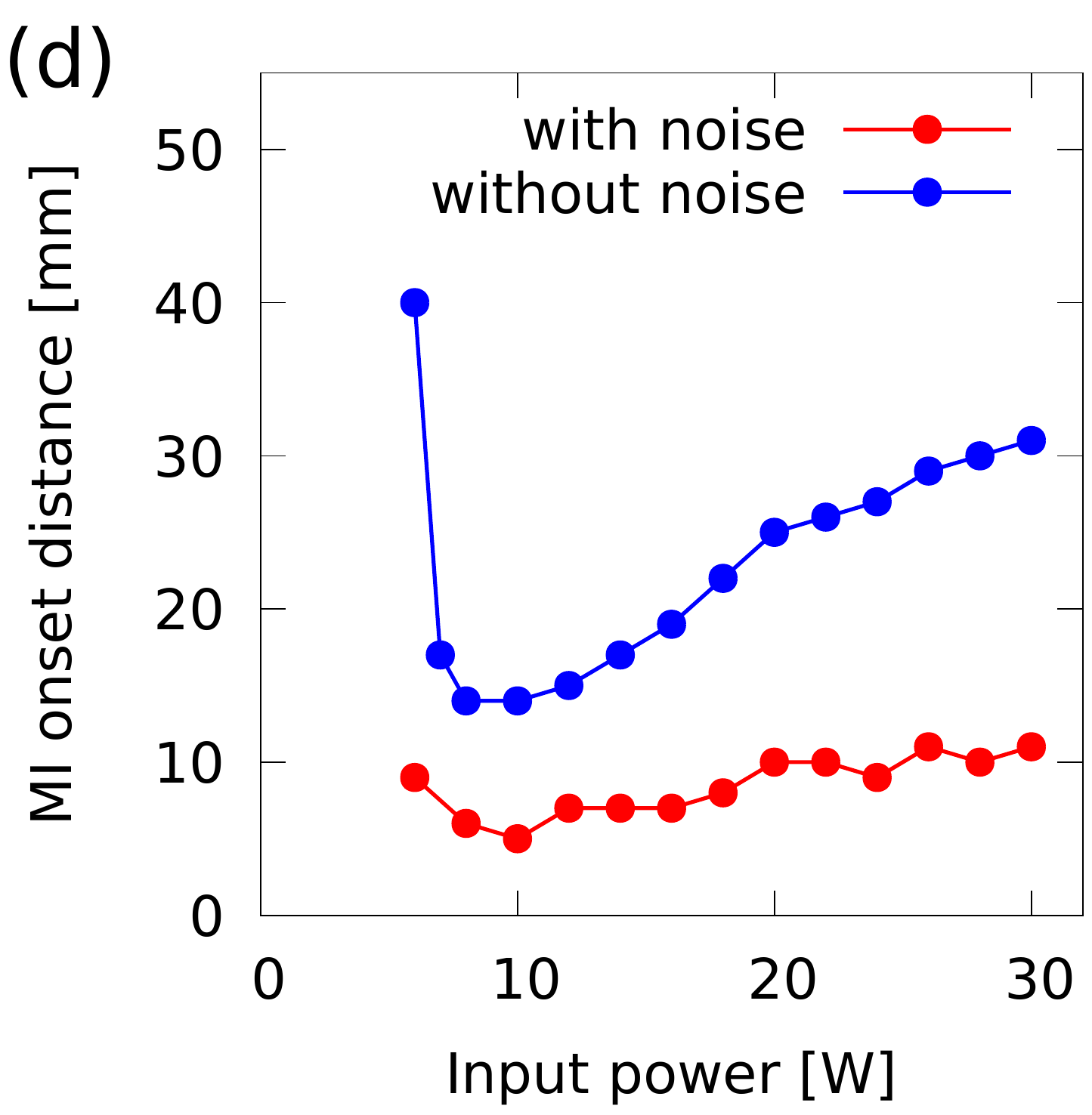}
	\caption{(a)--(c) Iso-intensity surfaces for a charge-two vortex beam with the initial radius $R = 20$~$\mu$m and powers 6~W (a), 8~W (stable vortex) (b), and 18~W (c), for the input without the noise. Propagation losses are neglected. (d)~Dependence of the distance at which the MI onsets as a function of the power of the vortex for input beams with (red) and without (blue) noise.}
	\label{fig:onset}
\end{figure}

First, we study the dependence of the distance at which the MI onsets as a function of the power of the vortex described by a Laguerre--Gaussian profile with a fixed charge $m=2$ and radius $R=20$~$\mu$m (radius is measured from the center of the vortex to the location of the peak intensity). Typical iso-intensity surfaces obtained during a 40-mm-long propagation are shown in Figs.~\ref{fig:onset}(a)--(c). \Cref{fig:onset}(b) shows the case where the input beam is the stable vortex solution. In the first $15$~mm of propagation, we can see a stationary propagation of the vortex, during which the transverse profile does not change. After that, the MI onsets and four solitons are generated and travel away from the vortex ring. At the power levels lower than that for the stable solution ($<8$~W), the diffraction of the vortex beam dominates. As it can be seen in \cref{fig:onset}(a), for the power level of $6$~W, in the first stage of propagation, the vortex beam diffracts but, the MI still manages to create the NB in a later stage of propagation. Below this value of power, MI is not observed and the vortex experiences only linear diffraction. At the power levels higher than that for the stable vortex soliton, the  MI onsets at a longer distance than for the stable vortex soliton because of the saturation of the nonlinear effects in the colloidal medium. In the regime close to saturation, the MI mediated growth of a perturbation with a given modulation amplitude will be slower than for the same perturbation at a power level below the saturation regime, due to different slopes of the intensity dependent refractive index, $n(I)$, curve in these two regimes. \Cref{fig:onset}(c) illustrates the evolution of the vortex beam at a high-power level, close to the nonlinearity saturation. At first, due to the excess of power, the radius of a vortex soliton oscillates, which resembles the behavior of higher-order solitons. Then, the MI creates a NB built of four separate solitons. This time however, due to the excess of power in each of the solitons and the presence of the OAM in the beam, the created solitons have an internal twisted structure, that reflects the presence of two closely spaced solitons rotating around a common center of mass. The plot in \cref{fig:onset}(d) summarizes the results of the study of the distance at which the MI onsets in function of the input power. The studies in which no noise was added to the input beam show a clear dependence of the MI onset distance on the input power. The MI onsets at the shortest distance for the power corresponding to the stable vortex solution. Deviation from this power in any direction results in increase of the MI onset distance either due to competition with diffraction (lower powers) or saturation effects (higher powers). Adding $5\%$ of white noise to the input beam accelerates the MI mediated formation of NBs and leads to a weaker dependence of the MI onset distance on power. Nonetheless, the character of this dependence is the same as in the noiseless case, as it can be seen in \cref{fig:onset}(d).

\begin{figure}[!t]
	\includegraphics[width = 0.49\columnwidth, clip = true, trim = {60 0 60 50}]{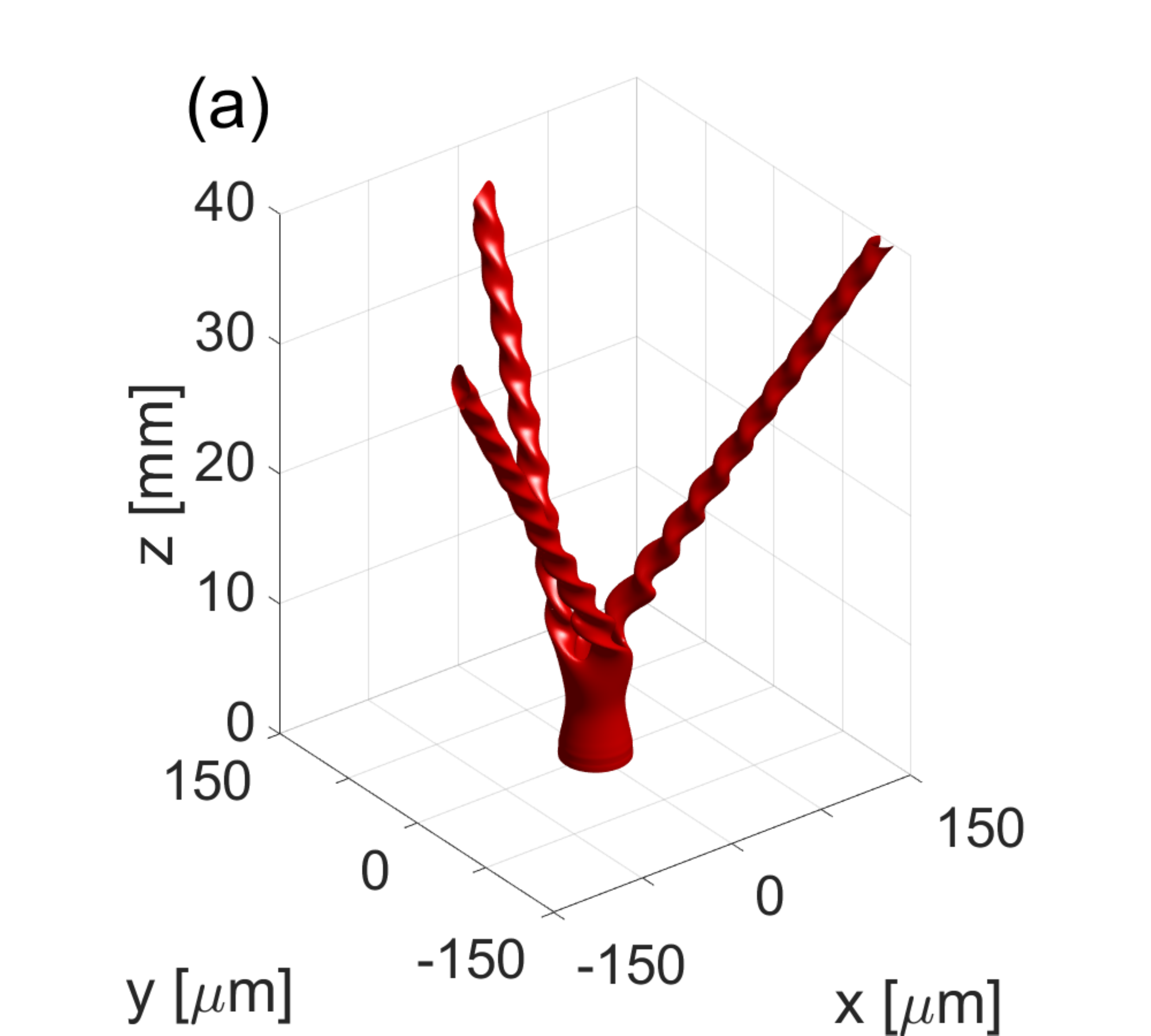}
	\includegraphics[width = 0.49\columnwidth, clip = true, trim = {60 0 60 50}]{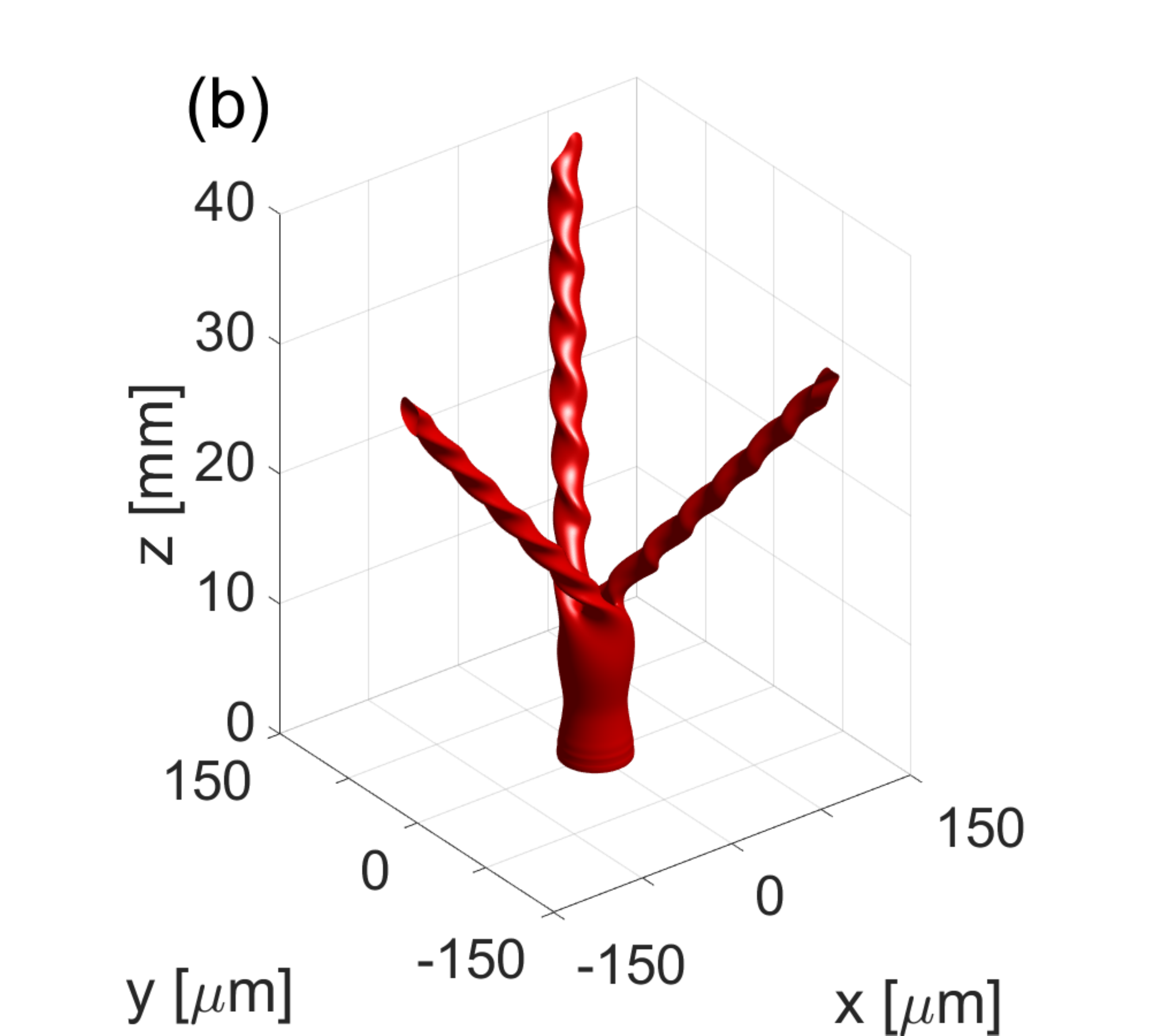}
	\includegraphics[width = 0.49\columnwidth, clip = true, trim = {60 0 60 220}]{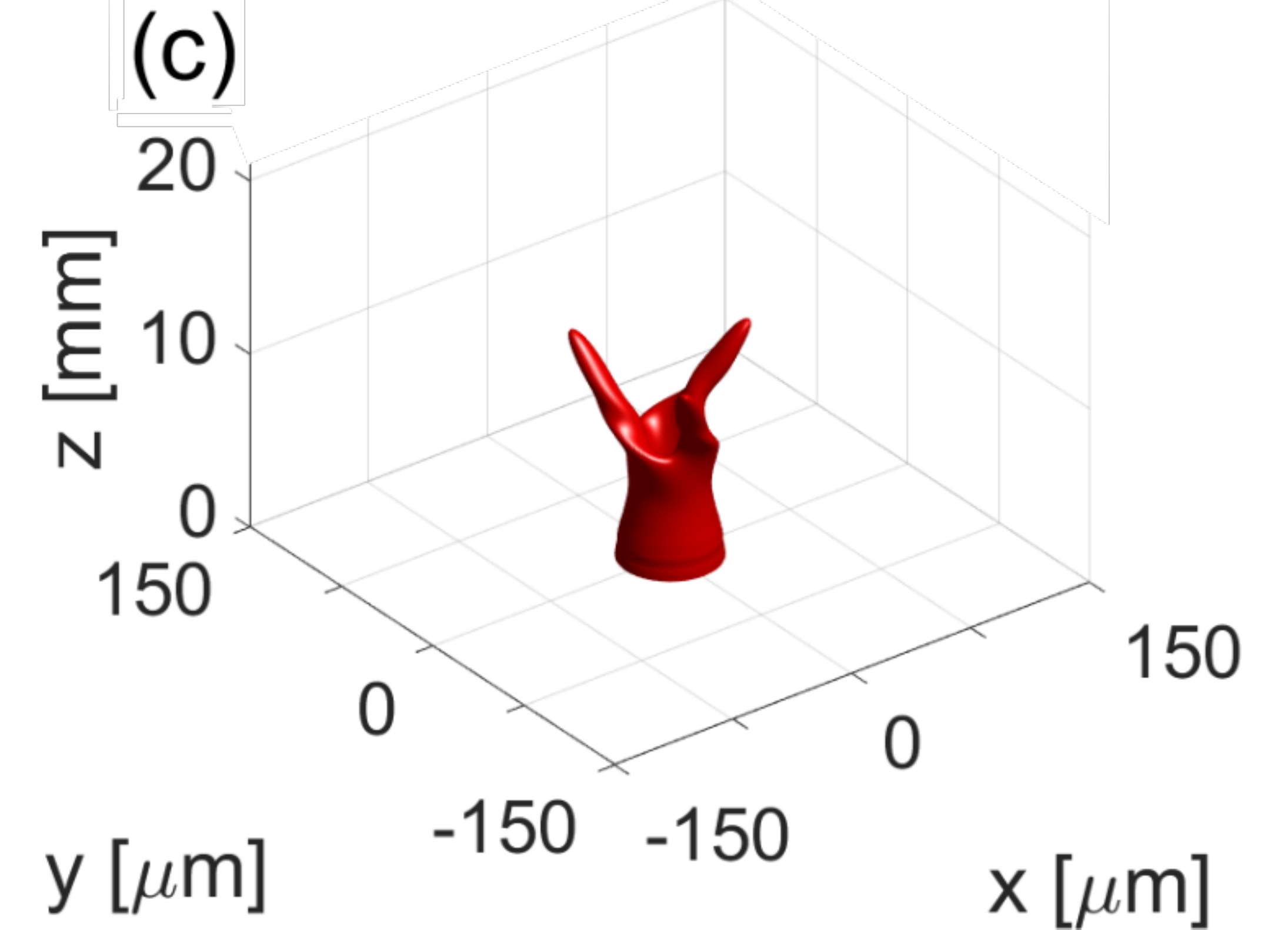}
	\includegraphics[width = 0.49\columnwidth, clip = true, trim = {60 0 60 220}]{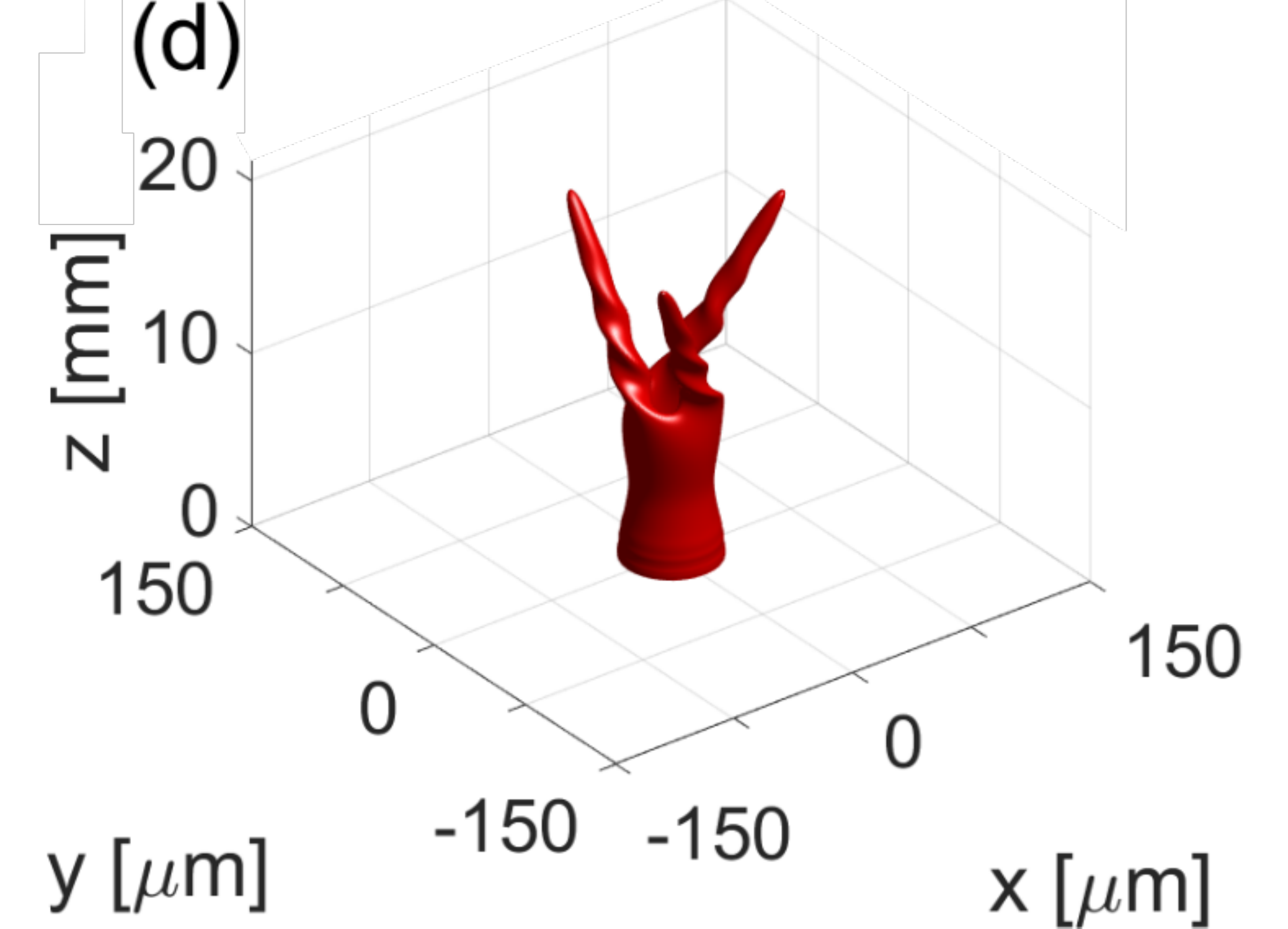}
	\caption{Iso-intensity surfaces for a charge-two vortex beams with radius $R = 20$~$\mu$m (same as is \cref{fig:onset}) for input powers 20~W (a), (c) and 30~W (b), (d) for the case with $5\%$ of noise added to the input. Panels (a) and (b) show the propagation in the medium without losses, while panels (c) and (d) show the result of simulations where losses are included.}
	\label{fig:loss}
\end{figure}

The presence of losses in the system prevents the NB formation at low-power levels but does not dramatically change the MI onset distance for high-power levels. In \cref{fig:loss}, we present a comparison of iso-intensity surfaces for the high-power beams propagating in systems with and without loss. Loss strongly limits the propagation distance of the solitons created via MI. Increase of the input power results in increased propagation distance of the created solitons due to the nonlinear self-induced transparency in the CS studied here~\cite{Man13}. With the increase of power, the loss experienced by the beam decreases, as the scattering particles with negative polarizability are expelled from the high-intensity region.

\begin{figure}[!t]
	\includegraphics[width = 0.49\columnwidth, clip = true, trim = {0 0 0 0}]{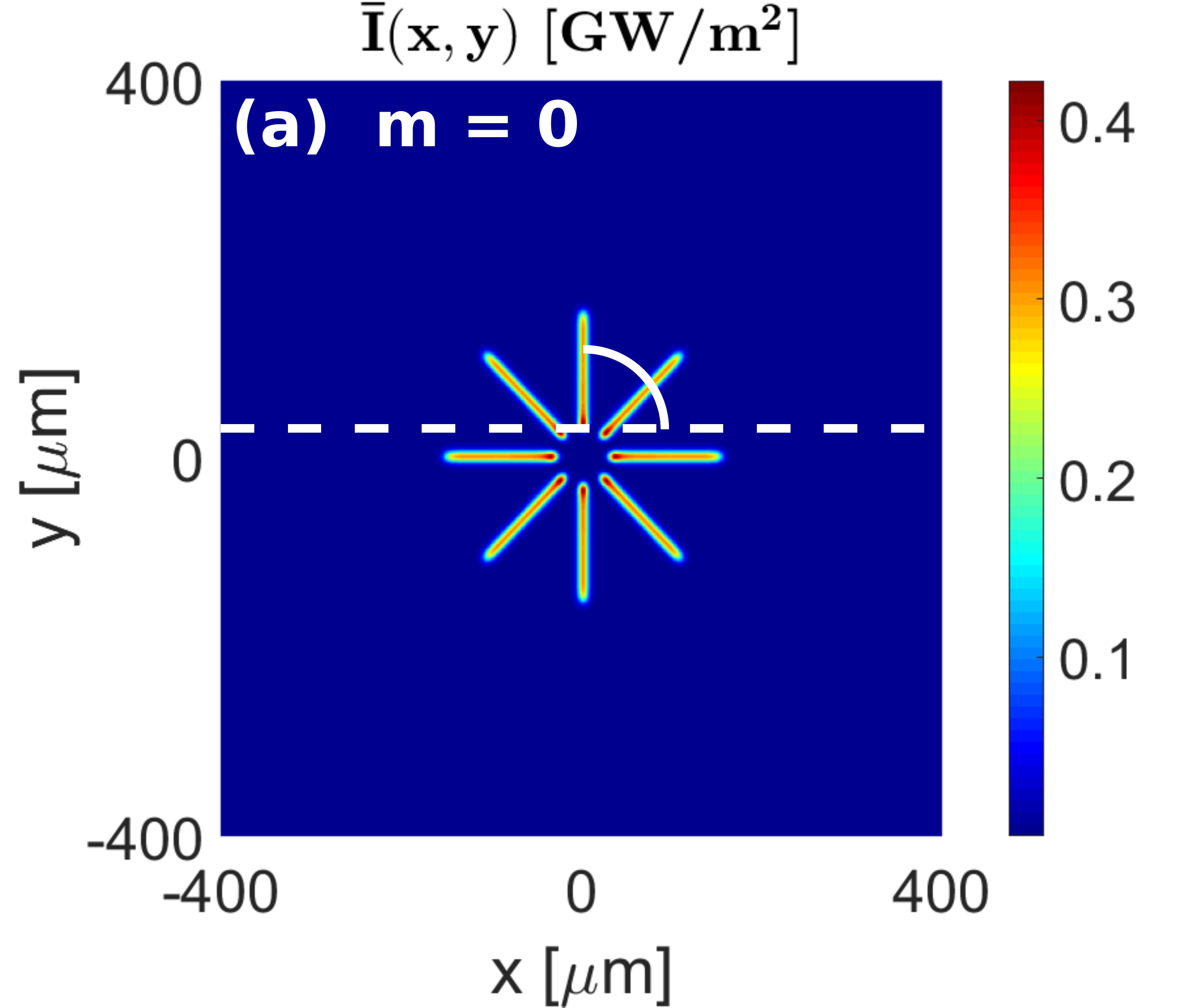}
	\includegraphics[width = 0.49\columnwidth, clip = true, trim = {0 0 0 0}]{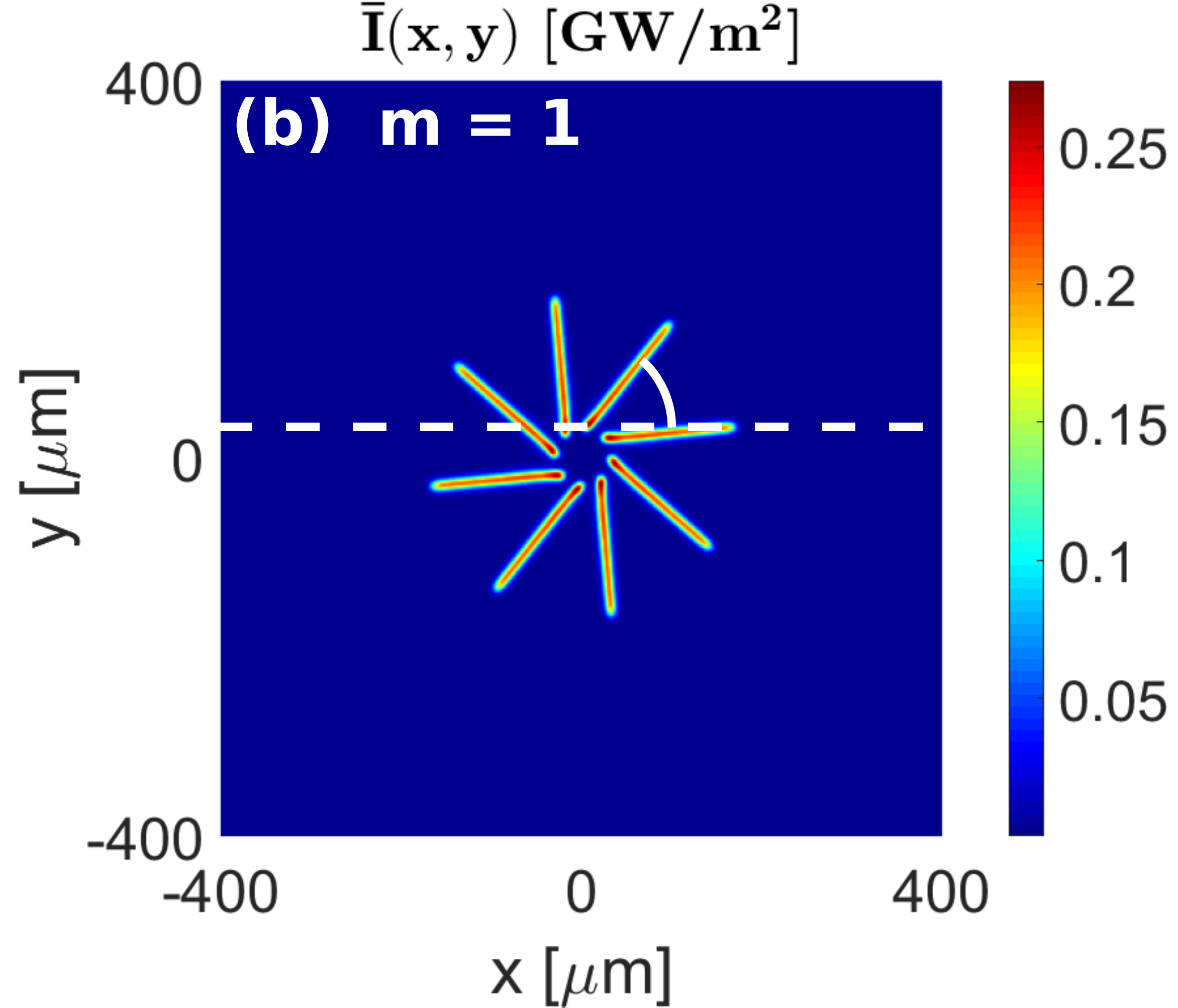}
	\includegraphics[width = 0.49\columnwidth, clip = true, trim = {0 0 0 0}]{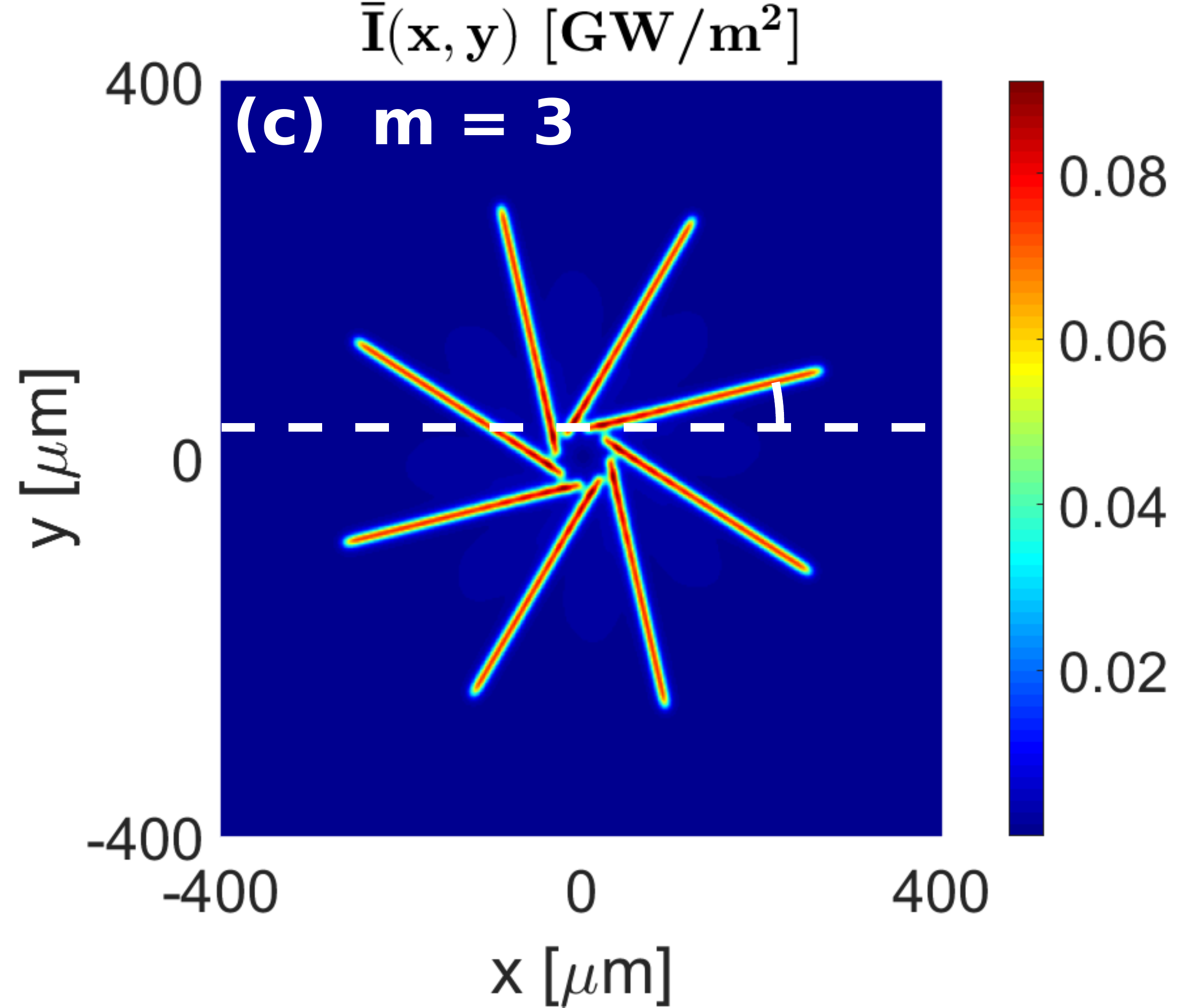}
	\includegraphics[width = 0.49\columnwidth, clip = true, trim = {0 0 0 0}]{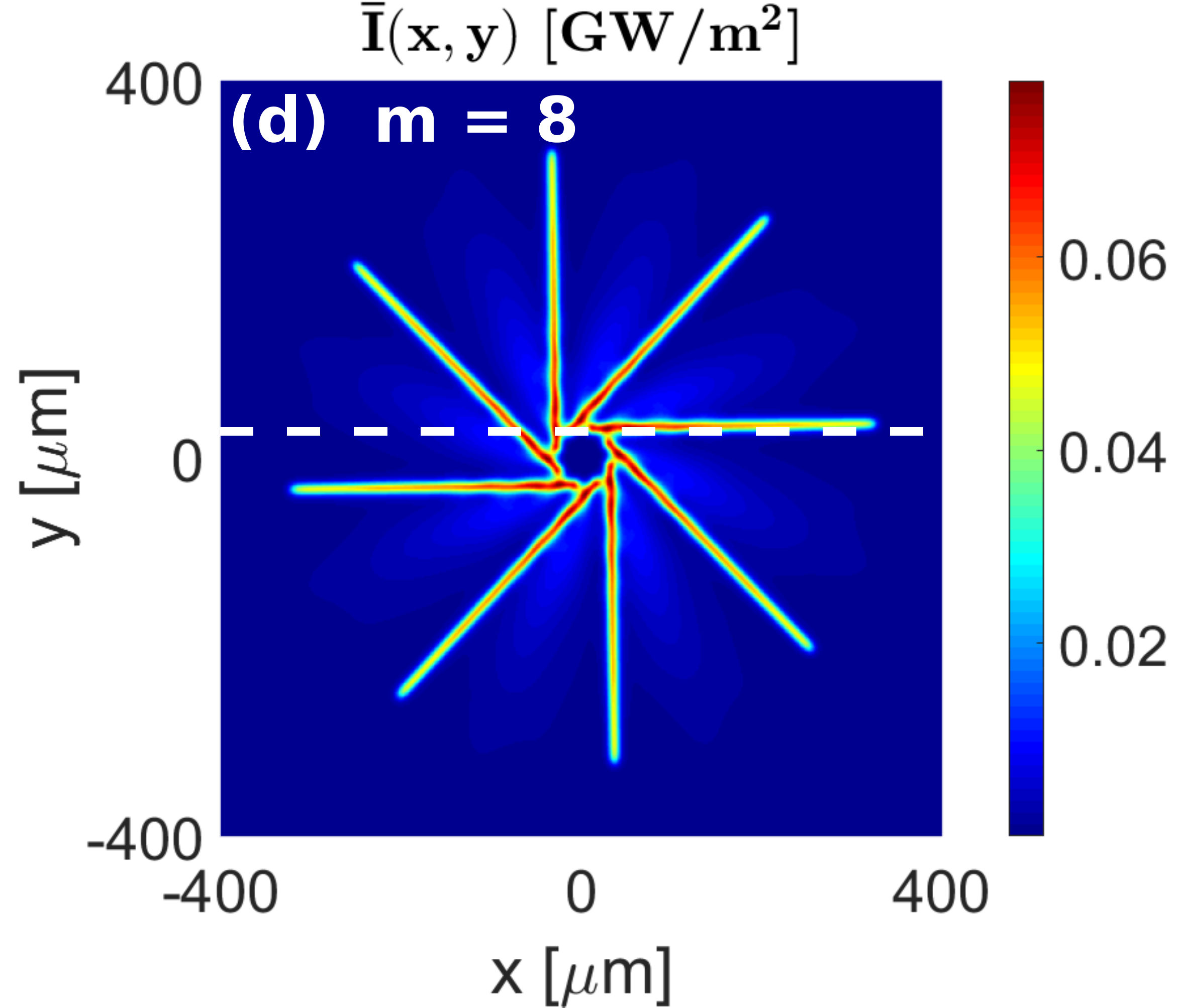}
	\includegraphics[width = 0.6\columnwidth, clip = true, trim = {0 0 0 0}]{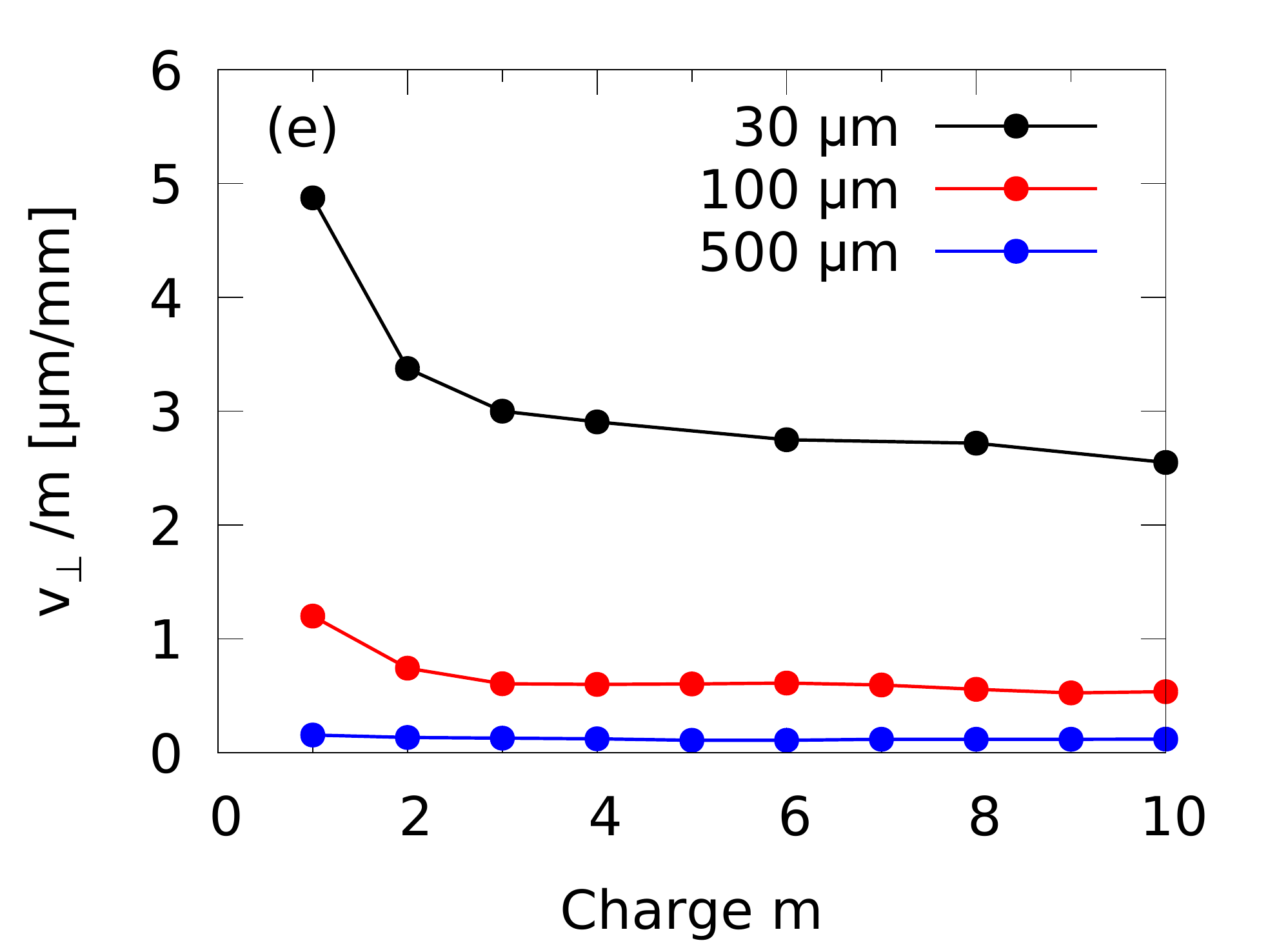}
	\caption{(a)--(d) Intensity maps averaged over the propagation distance [$\bar{I}(x,y) = 1/z_{\textrm{max}}\int_{0}^{z_{\textrm{max}}} I(x,y,z) \, \mathrm{d}z$] ($z_{\textrm{max}} = 40$~mm) for NBs with radius $R=30$~$\mu$m, eight maxima (solitons), input power 20~W, carrying various orbital angular momenta: $m=0$ (a), $m=1$ (b), $m=3$ (c), and $m=8$ ($z_{\textrm{max}} = 20$~mm)~(d). The white dashed line is tangent to the NB ring at the point located at the uppermost maximum for $z=0$. (e) Transverse escape velocity of the solitons normalized by the beam OAM ($v_{\perp}/m$) for various initial vortex radii: $30$~$\mu$m (black), $100$~$\mu$m (red), and $500$~$\mu$m (blue).} 
	\label{fig:escape}
\end{figure}

In the second part of this work, we investigate the trajectories and escape velocities of the solitons forming the NBs. Theoretical models predicted that the escape velocity of a soliton $v_{\perp}$, defined as the distance traveled in the transverse plane ($x$--$y$) divided by the distance traveled along the propagation axis ($z$), is proportional to the absolute value of the charge of the vortex and inversely proportional to the input vortex radius: $v_{\perp} \propto |m|/R$~\cite{Skryabin98,Soljacic00}. This prediction is valid for vortex beams with radius $R$ much larger than the width of the vortex donut [defined by the parameter $\sigma$ in \cref{eq:vortex-sol}]~\cite{Soljacic00}.  Additionally, the trajectories reported up to now were always tangent to the NB ring~\cite{Firth97,Skryabin98}. Here, we analyze the validity of this analytical prediction for various radii of the vortex beam, and show that the trajectories can create an arbitrary angle with the line tangent to the NB ring.

Depending on the radius, power level, and the charge of a vortex, the MI leads to formation of NBs with a different number of maxima~\cite{Skryabin98,Vinçotte06,Vuong06,Silahli:15}. Here, we want to analyze a system where we can control each of the beam parameters independently. To achieve that, instead of analyzing the MI-mediated breakup of the input vortex, we start from a NB with a predetermined number of maxima and fixed charge, radius, and input power. Such NBs can be generated using interference of two vortices with different charges but the same radii~\cite{Soljacic98,Soljacic00,Soljacic01}. The input profile with $N=8$ maxima and charge $m$ is described in cylindrical coordinates ($r$, $\theta$, $z$) by 
\begin{align}
\label{eq:vortex-sol}
\phi(r,\theta,z=0)   = f(r) \left[e^{i\left(m+\frac{N}{2} \right)\theta } + e^{i\left(m-\frac{N}{2} \right)\theta }\right],
\end{align} 
and the spatial profile $f(r)  = \cosh^{-1}[(r-R)/\sigma]$ closely resembles the vortex donut profile. We have chosen the width of the donut beam to be $\sigma = 0.2R$.

Figures~\ref{fig:escape}(a)--(d) show the $z$-averaged intensity maps $\bar{I}(x,y) = 1/z_{\textrm{max}}\int_{0}^{z_{\textrm{max}}} I(x,y,z) \, \mathrm{d}z$ that illustrate the trajectories of the soliton beams during the escape from the initial NB ring with the radius $R = 30$~$\mu$m. The angle between the trajectory of the uppermost soliton and the line tangent to the NB ring at $z=0$ is shown in the figures. We see that for charge $m=0$, the solitons propagate in the direction perpendicular to the NB ring (right angle between the beam trajectory and the tangent line). This movement is caused by the diffraction of the NB. As the charge of the NB increases, the angle between the beam trajectory and the tangent line decreases. The velocity of the solitons has now two components: the component perpendicular to the tangent line stemming from the beam diffraction and the component parallel to this line originating from the OAM carried by the beam. For high charges of the beam [e.g., $m=8$ shown in \cref{fig:escape}(d)], the contribution of diffraction is negligible compared to the OAM induced velocity. We have performed similar studies for NBs with radii $R = 100$~$\mu$m and $R = 500$~$\mu$m, and the results are summarized in \cref{fig:escape}(e). As expected, for beams with larger radius, the contribution of diffraction is smaller and it is visible only for low charges~$m$ of the beam. For radius $R= 500$~$\mu$m the contribution of diffraction is negligible for the whole range of charges studied and the theoretically predicted relation $v_{\perp}/|m| = \mathrm{const}(R)$ is fulfilled. 

\begin{figure}[!b]
	\includegraphics[width = 0.45\columnwidth, clip = true, trim = {160 0 220 0}]{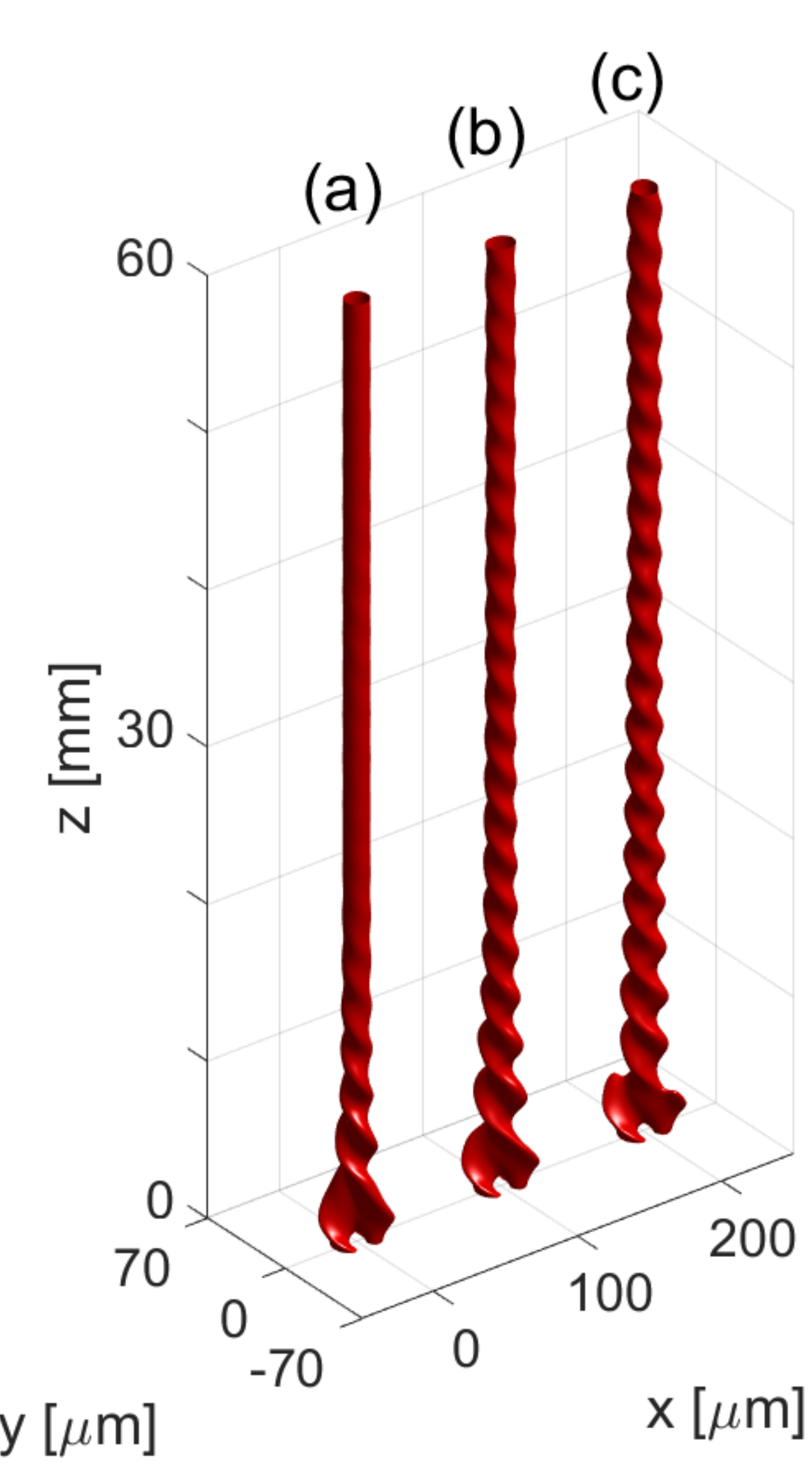}
	\includegraphics[width = 0.53\columnwidth, clip = true, trim = {125 0 200 0}]{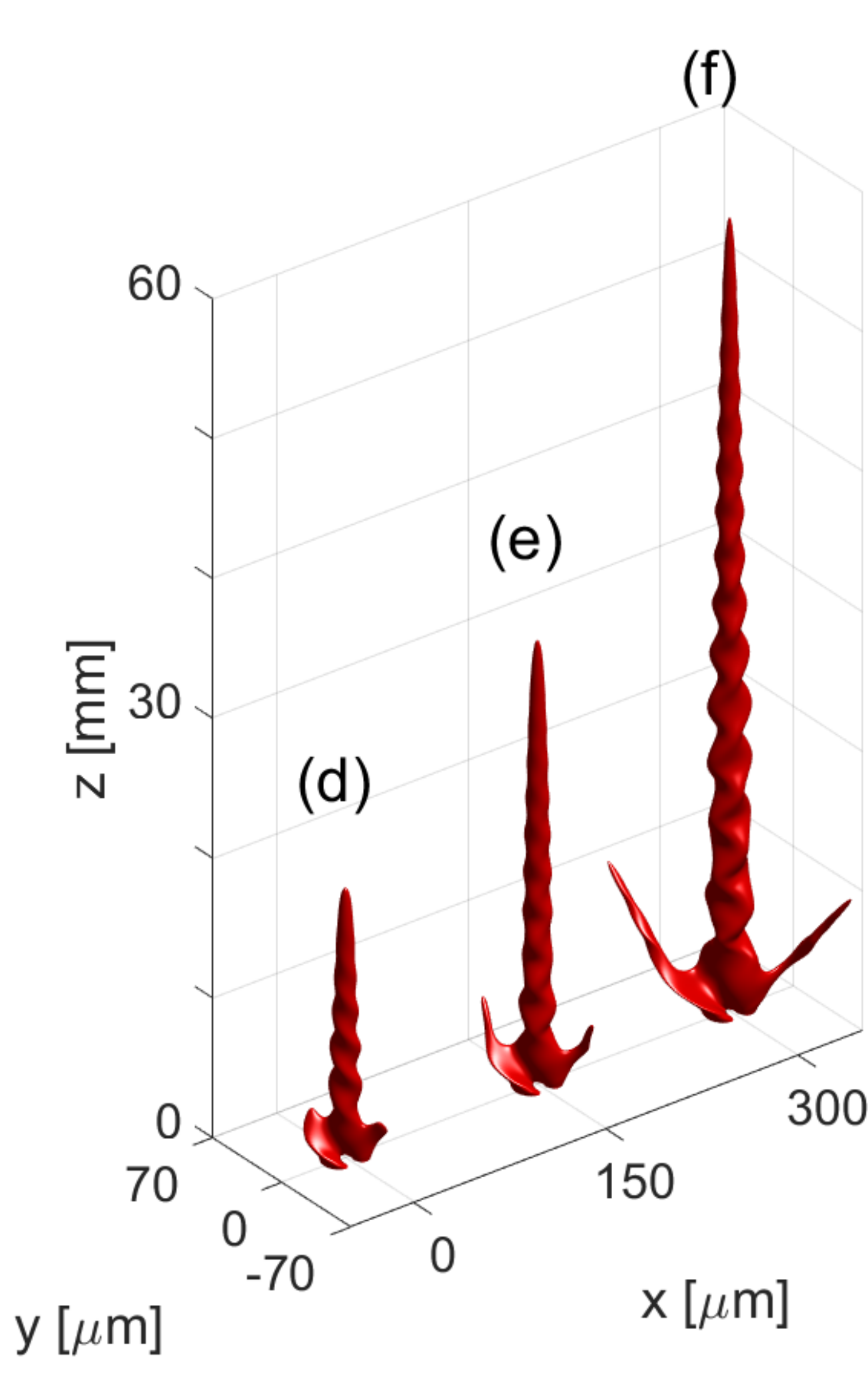}
	\caption{Twisted solitons generated in systems without loss (a)--(c) and with loss (d)--(f). The input is a NB with $m=1$, $R=10$~$\mu$m, and two maxima. The input power levels are: 4~W (a), 6~W (b), 8~W (c), 12~W (d), 20~W~(e), and 40~W~(f).}
	\label{fig:twisted}
\end{figure}

In the previous paragraphs, we have studied situations in which the solitons building the NB were traveling away from the vortex ring during their propagation. Different scenarios are also possible, in which the solitons forming the NB interact with each other. For the case of two interacting solitons ($N=2$) spiraling and twisted beams can be formed. 
Spiraling solitons are build of two separate beams rotating around each other as they propagate. Generation of stable spiraling solitons was theoretically and experimentally demonstrated in materials with nonlocal nonlinearities~\cite{Tikhonenko:95,Tikhonenko96,Buccoliero07,Buccoliero:08,Liang13,Poladian91,Shih97,Belic99,Buryak99,Lem-Carrillo14}. Since the nonlinearity in our model is assumed to be local and instantaneous, the spiraling beams are unstable in such a medium~\cite{Steblina:98}. Above the critical radius, the two beams escape from each other; and below this radius, the beams fuse. This latter phenomena generates twisted beams: corkscrew-like beams where two tightly spaced solitons rotate around the common center of mass. \Cref{fig:twisted} shows the propagation of such beams in lossless and lossy media for various input-power levels. In the lossless case, at low power, the twisted beam is formed at the beginning of the propagation but after approximately $20$~mm the twist is no longer visible and the beam propagates as a regular soliton. The excess of energy, that is a result of the fusion of two beams, is expelled from the beam and only a single stable soliton remains. For higher input energies, the twisted nature of the beam is preserved for a longer propagation distance, as it takes more time to expel the energy excess. Increase of the energy does not allow for formation of solitons of the order higher than one due to the saturable nature of the nonlinearity of the studied CS. In the case of propagation in the lossy medium, the twisted beam maintains the twisted character and propagates for longer distance as the power is increased, due to the self-induced transparency effect in the negative polarizability CSs~\cite{Man13}. 

\begin{figure}[!b]
	\includegraphics[width = 0.7\columnwidth, clip = true, trim = {0 0 0 0}]{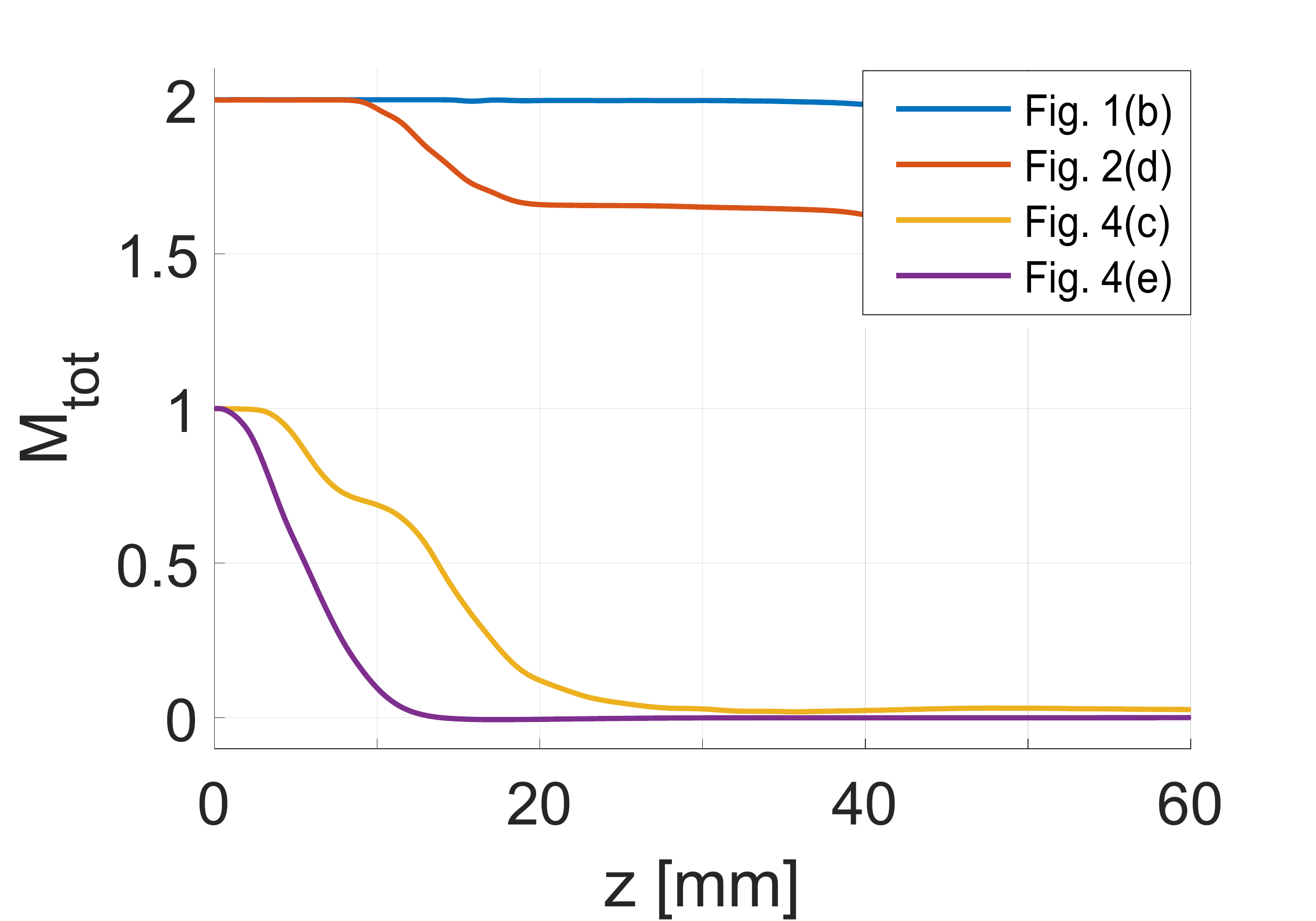}
	\caption{The total OAM as a function of propagation distance for necklace beams (blue and red curves) and twisted beams (yellow and violet curves) shown in the plots indicated in the legend. OAM conservation during propagation in colloidal media without loss (blue and yellow curves), and with loss (red and violet curves) is analyzed.}
	\label{fig:OAM}
\end{figure}

Finally, we have studied the conservation of the OAM of the NBs and the twisted beams. The total OAM is calculated as
\begin{equation}
M_{\textrm{tot}} =  \frac{\frac{i}{2} \iint \mathbf{r} \times \left( \phi \nabla_{\perp} \phi^* - \phi^* \nabla_{\perp} \phi \right) \, \mathrm{d}x \mathrm{d}y}{\iint |\phi|^2 \, \mathrm{d}x \mathrm{d}y}, \nonumber
\end{equation}
where $\mathbf{r} = [x,y]$ is a vector of position in the transverse plane and $\nabla_{\perp} = \left[\frac{\partial}{\partial x}, \frac{\partial }{\partial y}\right]$.
\Cref{fig:OAM} presents plots of the total OAM as a function of the propagation distance for four beams studied above: two NBs and two twisted beams. Blue and red curves show the $M_{\textrm{tot}}(z)$ for NBs without and with losses, respectively. In the lossless case, for which the iso-intensity surface is shown in \cref{fig:onset}(b), the OAM is conserved both when the vortex beam propagates and transforms into the NB, and during the propagation of the solitons. In the case of the NB generation in a lossy medium [iso-intensity surface shown in \cref{fig:loss}(d)], the OAM is not fully conserved. It is conserved during the propagation of the vortex beam, but when the NB is formed, part of the OAM is lost. When the solitons lose energy and diffract, the OAM does not decrease any more. The loss of the OAM in the middle stage of the propagation is caused by fact that the three solitons forming the NB do not have equal energies, as they grow from random noise due to the MI. In our model, loss is intensity dependent, and the soliton with the lower power loses the energy faster than solitons with higher power. This results in non-symmetric distribution of the energy and leads to changes in the OAM carried by the NB. When beam intensities decrease, all the beams decay with the same rate (corresponding to the loss level in the linear regime), the OAM of the linearly diffracting field becomes conserved again.

For the twisted beams resulting from the fusion of two solitons, the OAM is not conserved as shown by the yellow and violet curves \cref{fig:OAM}. Here, the OAM is expelled from the twisted high-intensity beam and it is carried by the low-intensity field (not visible in iso-intensity plots) away from the central beam. In the lossless system, for which the iso-intensity surface is shown in \cref{fig:twisted}(c), the low-intensity field carrying the OAM is absorbed when it reaches the simulation domain boundaries. 
Surprisingly, the twisted character of the beam is present, despite the fact that the beam does no longer carry the OAM. In the system with losses [see \cref{fig:twisted}(e) and the corresponding violet curve in \cref{fig:OAM}], the low-intensity field that carries the OAM is rapidly absorbed. 
The OAM in this system decreases to zero in the first $10$~mm of propagation.

In conclusion, we investigated the dynamics of vortices in nonlinear colloidal suspensions and showed that the distance at which the modulation instability transforms a vortex beam into a necklace beam depends on the input power, and that the minimum distance is achieved for the power corresponding to the stationary vortex soliton. Moreover, we demonstrated the influence of diffraction on the trajectories of necklace beams and the value of the escape velocity. The solitons forming the necklace beam were shown to follow linear trajectories that are not necessarily tangent to the necklace beam ring, and the angle of escape can be controlled by the interplay of the necklace beam radius and charge. We showed the possibility to form twisted solitons from necklace beams with only two maxima and analyzed the influence of losses on the propagation of such types of beams. Finally, we investigated the conservation of orbital angular momentum in necklace beams and twisted beams. These results may be of interest for the formation of complex photonic structures in liquids as well as in the broader field of light filamentation.


\begin{thebibliography}{45}%
	\makeatletter
	\providecommand \@ifxundefined [1]{%
		\@ifx{#1\undefined}
	}%
	\providecommand \@ifnum [1]{%
		\ifnum #1\expandafter \@firstoftwo
		\else \expandafter \@secondoftwo
		\fi
	}%
	\providecommand \@ifx [1]{%
		\ifx #1\expandafter \@firstoftwo
		\else \expandafter \@secondoftwo
		\fi
	}%
	\providecommand \natexlab [1]{#1}%
	\providecommand \enquote  [1]{``#1''}%
	\providecommand \bibnamefont  [1]{#1}%
	\providecommand \bibfnamefont [1]{#1}%
	\providecommand \citenamefont [1]{#1}%
	\providecommand \href@noop [0]{\@secondoftwo}%
	\providecommand \href [0]{\begingroup \@sanitize@url \@href}%
	\providecommand \@href[1]{\@@startlink{#1}\@@href}%
	\providecommand \@@href[1]{\endgroup#1\@@endlink}%
	\providecommand \@sanitize@url [0]{\catcode `\\12\catcode `\$12\catcode
		`\&12\catcode `\#12\catcode `\^12\catcode `\_12\catcode `\%12\relax}%
	\providecommand \@@startlink[1]{}%
	\providecommand \@@endlink[0]{}%
	\providecommand \url  [0]{\begingroup\@sanitize@url \@url }%
	\providecommand \@url [1]{\endgroup\@href {#1}{\urlprefix }}%
	\providecommand \urlprefix  [0]{URL }%
	\providecommand \Eprint [0]{\href }%
	\providecommand \doibase [0]{http://dx.doi.org/}%
	\providecommand \selectlanguage [0]{\@gobble}%
	\providecommand \bibinfo  [0]{\@secondoftwo}%
	\providecommand \bibfield  [0]{\@secondoftwo}%
	\providecommand \translation [1]{[#1]}%
	\providecommand \BibitemOpen [0]{}%
	\providecommand \bibitemStop [0]{}%
	\providecommand \bibitemNoStop [0]{.\EOS\space}%
	\providecommand \EOS [0]{\spacefactor3000\relax}%
	\providecommand \BibitemShut  [1]{\csname bibitem#1\endcsname}%
	\let\auto@bib@innerbib\@empty
	\bibitem [{\citenamefont {Allen}\ \emph {et~al.}(1992)\citenamefont {Allen},
		\citenamefont {Beijersbergen}, \citenamefont {Spreeuw},\ and\ \citenamefont
		{Woerdman}}]{Allen92}%
	\BibitemOpen
	\bibfield  {author} {\bibinfo {author} {\bibfnamefont {L.}~\bibnamefont
			{Allen}}, \bibinfo {author} {\bibfnamefont {M.~W.}\ \bibnamefont
			{Beijersbergen}}, \bibinfo {author} {\bibfnamefont {R.~J.~C.}\ \bibnamefont
			{Spreeuw}}, \ and\ \bibinfo {author} {\bibfnamefont {J.~P.}\ \bibnamefont
			{Woerdman}},\ }\href {\doibase 10.1103/PhysRevA.45.8185} {\bibfield
		{journal} {\bibinfo  {journal} {Phys. Rev. A}\ }\textbf {\bibinfo {volume}
			{45}},\ \bibinfo {pages} {8185} (\bibinfo {year} {1992})}\BibitemShut
	{NoStop}%
	\bibitem [{\citenamefont {Ch\'avez-Cerda}\ \emph {et~al.}(1996)\citenamefont
		{Ch\'avez-Cerda}, \citenamefont {McDonald},\ and\ \citenamefont
		{New}}]{CHAVEZCERDA1996}%
	\BibitemOpen
	\bibfield  {author} {\bibinfo {author} {\bibfnamefont {S.}~\bibnamefont
			{Ch\'avez-Cerda}}, \bibinfo {author} {\bibfnamefont {G.}~\bibnamefont
			{McDonald}}, \ and\ \bibinfo {author} {\bibfnamefont {G.}~\bibnamefont
			{New}},\ }\href {\doibase 10.1016/0030-4018(95)00538-2} {\bibfield  {journal}
		{\bibinfo  {journal} {Opt. Commun.}\ }\textbf {\bibinfo {volume} {123}},\
		\bibinfo {pages} {225 } (\bibinfo {year} {1996})}\BibitemShut {NoStop}%
	\bibitem [{\citenamefont {Schechner}\ \emph {et~al.}(1996)\citenamefont
		{Schechner}, \citenamefont {Piestun},\ and\ \citenamefont
		{Shamir}}]{Schechner96}%
	\BibitemOpen
	\bibfield  {author} {\bibinfo {author} {\bibfnamefont {Y.~Y.}\ \bibnamefont
			{Schechner}}, \bibinfo {author} {\bibfnamefont {R.}~\bibnamefont {Piestun}},
		\ and\ \bibinfo {author} {\bibfnamefont {J.}~\bibnamefont {Shamir}},\ }\href
	{\doibase 10.1103/PhysRevE.54.R50} {\bibfield  {journal} {\bibinfo  {journal}
			{Phys. Rev. E}\ }\textbf {\bibinfo {volume} {54}},\ \bibinfo {pages} {R50}
		(\bibinfo {year} {1996})}\BibitemShut {NoStop}%
	\bibitem [{\citenamefont {Abramochkin}\ \emph {et~al.}(1997)\citenamefont
		{Abramochkin}, \citenamefont {Losevsky},\ and\ \citenamefont
		{Volostnikov}}]{ABRAMOCHKIN1997}%
	\BibitemOpen
	\bibfield  {author} {\bibinfo {author} {\bibfnamefont {E.}~\bibnamefont
			{Abramochkin}}, \bibinfo {author} {\bibfnamefont {N.}~\bibnamefont
			{Losevsky}}, \ and\ \bibinfo {author} {\bibfnamefont {V.}~\bibnamefont
			{Volostnikov}},\ }\href {\doibase 10.1016/S0030-4018(97)00215-0} {\bibfield
		{journal} {\bibinfo  {journal} {Opt. Commun.}\ }\textbf {\bibinfo {volume}
			{141}},\ \bibinfo {pages} {59 } (\bibinfo {year} {1997})}\BibitemShut
	{NoStop}%
	\bibitem [{\citenamefont {Tervo}\ and\ \citenamefont
		{Turunen}(2001)}]{Tervo:01}%
	\BibitemOpen
	\bibfield  {author} {\bibinfo {author} {\bibfnamefont {J.}~\bibnamefont
			{Tervo}}\ and\ \bibinfo {author} {\bibfnamefont {J.}~\bibnamefont
			{Turunen}},\ }\href {\doibase 10.1364/OE.9.000009} {\bibfield  {journal}
		{\bibinfo  {journal} {Opt. Express}\ }\textbf {\bibinfo {volume} {9}},\
		\bibinfo {pages} {9} (\bibinfo {year} {2001})}\BibitemShut {NoStop}%
	\bibitem [{\citenamefont {Schulze}\ \emph {et~al.}(2015)\citenamefont
		{Schulze}, \citenamefont {Roux}, \citenamefont {Dudley}, \citenamefont {Rop},
		\citenamefont {Duparr\'e},\ and\ \citenamefont {Forbes}}]{Schulze15}%
	\BibitemOpen
	\bibfield  {author} {\bibinfo {author} {\bibfnamefont {C.}~\bibnamefont
			{Schulze}}, \bibinfo {author} {\bibfnamefont {F.~S.}\ \bibnamefont {Roux}},
		\bibinfo {author} {\bibfnamefont {A.}~\bibnamefont {Dudley}}, \bibinfo
		{author} {\bibfnamefont {R.}~\bibnamefont {Rop}}, \bibinfo {author}
		{\bibfnamefont {M.}~\bibnamefont {Duparr\'e}}, \ and\ \bibinfo {author}
		{\bibfnamefont {A.}~\bibnamefont {Forbes}},\ }\href {\doibase
		10.1103/PhysRevA.91.043821} {\bibfield  {journal} {\bibinfo  {journal} {Phys.
				Rev. A}\ }\textbf {\bibinfo {volume} {91}},\ \bibinfo {pages} {043821}
		(\bibinfo {year} {2015})}\BibitemShut {NoStop}%
	\bibitem [{\citenamefont {Wei-Ping}\ \emph {et~al.}(2010)\citenamefont
		{Wei-Ping}, \citenamefont {Belić}, \citenamefont {Ting-Wen},\ and\
		\citenamefont {Li-Yang}}]{Wei-Ping10}%
	\BibitemOpen
	\bibfield  {author} {\bibinfo {author} {\bibfnamefont {Z.}~\bibnamefont
			{Wei-Ping}}, \bibinfo {author} {\bibfnamefont {M.}~\bibnamefont {Belić}},
		\bibinfo {author} {\bibfnamefont {H.}~\bibnamefont {Ting-Wen}}, \ and\
		\bibinfo {author} {\bibfnamefont {W.}~\bibnamefont {Li-Yang}},\ }\href
	{http://stacks.iop.org/0253-6102/53/i=4/a=30} {\bibfield  {journal} {\bibinfo
			{journal} {Commun. Theor. Phys.}\ }\textbf {\bibinfo {volume} {53}},\
		\bibinfo {pages} {749} (\bibinfo {year} {2010})}\BibitemShut {NoStop}%
	\bibitem [{\citenamefont {Tikhonenko}\ \emph {et~al.}(1995)\citenamefont
		{Tikhonenko}, \citenamefont {Christou},\ and\ \citenamefont
		{Luther-Daves}}]{Tikhonenko:95}%
	\BibitemOpen
	\bibfield  {author} {\bibinfo {author} {\bibfnamefont {V.}~\bibnamefont
			{Tikhonenko}}, \bibinfo {author} {\bibfnamefont {J.}~\bibnamefont
			{Christou}}, \ and\ \bibinfo {author} {\bibfnamefont {B.}~\bibnamefont
			{Luther-Daves}},\ }\href {\doibase 10.1364/JOSAB.12.002046} {\bibfield
		{journal} {\bibinfo  {journal} {J. Opt. Soc. Am. B}\ }\textbf {\bibinfo
			{volume} {12}},\ \bibinfo {pages} {2046} (\bibinfo {year}
		{1995})}\BibitemShut {NoStop}%
	\bibitem [{\citenamefont {Tikhonenko}\ \emph {et~al.}(1996)\citenamefont
		{Tikhonenko}, \citenamefont {Christou},\ and\ \citenamefont
		{Luther-Davies}}]{Tikhonenko96}%
	\BibitemOpen
	\bibfield  {author} {\bibinfo {author} {\bibfnamefont {V.}~\bibnamefont
			{Tikhonenko}}, \bibinfo {author} {\bibfnamefont {J.}~\bibnamefont
			{Christou}}, \ and\ \bibinfo {author} {\bibfnamefont {B.}~\bibnamefont
			{Luther-Davies}},\ }\href {\doibase 10.1103/PhysRevLett.76.2698} {\bibfield
		{journal} {\bibinfo  {journal} {Phys. Rev. Lett.}\ }\textbf {\bibinfo
			{volume} {76}},\ \bibinfo {pages} {2698} (\bibinfo {year}
		{1996})}\BibitemShut {NoStop}%
	\bibitem [{\citenamefont {Mihalache}\ \emph {et~al.}(2002)\citenamefont
		{Mihalache}, \citenamefont {Mazilu}, \citenamefont {Crasovan}, \citenamefont
		{Towers}, \citenamefont {Buryak}, \citenamefont {Malomed}, \citenamefont
		{Torner}, \citenamefont {Torres},\ and\ \citenamefont
		{Lederer}}]{Mihalache02}%
	\BibitemOpen
	\bibfield  {author} {\bibinfo {author} {\bibfnamefont {D.}~\bibnamefont
			{Mihalache}}, \bibinfo {author} {\bibfnamefont {D.}~\bibnamefont {Mazilu}},
		\bibinfo {author} {\bibfnamefont {L.-C.}\ \bibnamefont {Crasovan}}, \bibinfo
		{author} {\bibfnamefont {I.}~\bibnamefont {Towers}}, \bibinfo {author}
		{\bibfnamefont {A.~V.}\ \bibnamefont {Buryak}}, \bibinfo {author}
		{\bibfnamefont {B.~A.}\ \bibnamefont {Malomed}}, \bibinfo {author}
		{\bibfnamefont {L.}~\bibnamefont {Torner}}, \bibinfo {author} {\bibfnamefont
			{J.~P.}\ \bibnamefont {Torres}}, \ and\ \bibinfo {author} {\bibfnamefont
			{F.}~\bibnamefont {Lederer}},\ }\href {\doibase
		10.1103/PhysRevLett.88.073902} {\bibfield  {journal} {\bibinfo  {journal}
			{Phys. Rev. Lett.}\ }\textbf {\bibinfo {volume} {88}},\ \bibinfo {pages}
		{073902} (\bibinfo {year} {2002})}\BibitemShut {NoStop}%
	\bibitem [{\citenamefont {Buccoliero}\ \emph {et~al.}(2007)\citenamefont
		{Buccoliero}, \citenamefont {Lopez-Aguayo}, \citenamefont {Skupin},
		\citenamefont {Desyatnikov}, \citenamefont {Bang}, \citenamefont
		{Krolikowski},\ and\ \citenamefont {Kivshar}}]{Buccoliero07}%
	\BibitemOpen
	\bibfield  {author} {\bibinfo {author} {\bibfnamefont {D.}~\bibnamefont
			{Buccoliero}}, \bibinfo {author} {\bibfnamefont {S.}~\bibnamefont
			{Lopez-Aguayo}}, \bibinfo {author} {\bibfnamefont {S.}~\bibnamefont
			{Skupin}}, \bibinfo {author} {\bibfnamefont {A.~S.}\ \bibnamefont
			{Desyatnikov}}, \bibinfo {author} {\bibfnamefont {O.}~\bibnamefont {Bang}},
		\bibinfo {author} {\bibfnamefont {W.}~\bibnamefont {Krolikowski}}, \ and\
		\bibinfo {author} {\bibfnamefont {Y.~S.}\ \bibnamefont {Kivshar}},\ }\href
	{\doibase 10.1016/j.physb.2006.12.063} {\bibfield  {journal} {\bibinfo
			{journal} {Physica B: Condens. Matter}\ }\textbf {\bibinfo {volume} {394}},\
		\bibinfo {pages} {351 } (\bibinfo {year} {2007})}\BibitemShut {NoStop}%
	\bibitem [{\citenamefont {Buccoliero}\ \emph {et~al.}(2008)\citenamefont
		{Buccoliero}, \citenamefont {Desyatnikov}, \citenamefont {Krolikowski},\ and\
		\citenamefont {Kivshar}}]{Buccoliero:08}%
	\BibitemOpen
	\bibfield  {author} {\bibinfo {author} {\bibfnamefont {D.}~\bibnamefont
			{Buccoliero}}, \bibinfo {author} {\bibfnamefont {A.~S.}\ \bibnamefont
			{Desyatnikov}}, \bibinfo {author} {\bibfnamefont {W.}~\bibnamefont
			{Krolikowski}}, \ and\ \bibinfo {author} {\bibfnamefont {Y.~S.}\ \bibnamefont
			{Kivshar}},\ }\href {\doibase 10.1364/OL.33.000198} {\bibfield  {journal}
		{\bibinfo  {journal} {Opt. Lett.}\ }\textbf {\bibinfo {volume} {33}},\
		\bibinfo {pages} {198} (\bibinfo {year} {2008})}\BibitemShut {NoStop}%
	\bibitem [{\citenamefont {Liang}\ and\ \citenamefont {Guo}(2013)}]{Liang13}%
	\BibitemOpen
	\bibfield  {author} {\bibinfo {author} {\bibfnamefont {G.}~\bibnamefont
			{Liang}}\ and\ \bibinfo {author} {\bibfnamefont {Q.}~\bibnamefont {Guo}},\
	}\href {\doibase 10.1103/PhysRevA.88.043825} {\bibfield  {journal} {\bibinfo
			{journal} {Phys. Rev. A}\ }\textbf {\bibinfo {volume} {88}},\ \bibinfo
		{pages} {043825} (\bibinfo {year} {2013})}\BibitemShut {NoStop}%
	\bibitem [{\citenamefont {Wen}\ \emph {et~al.}(2016)\citenamefont {Wen},
		\citenamefont {Chen}, \citenamefont {Zhang}, \citenamefont {Chen},\ and\
		\citenamefont {Yu}}]{Wen16}%
	\BibitemOpen
	\bibfield  {author} {\bibinfo {author} {\bibfnamefont {Y.}~\bibnamefont
			{Wen}}, \bibinfo {author} {\bibfnamefont {Y.}~\bibnamefont {Chen}}, \bibinfo
		{author} {\bibfnamefont {Y.}~\bibnamefont {Zhang}}, \bibinfo {author}
		{\bibfnamefont {H.}~\bibnamefont {Chen}}, \ and\ \bibinfo {author}
		{\bibfnamefont {S.}~\bibnamefont {Yu}},\ }\href {\doibase
		10.1103/PhysRevA.94.013829} {\bibfield  {journal} {\bibinfo  {journal} {Phys.
				Rev. A}\ }\textbf {\bibinfo {volume} {94}},\ \bibinfo {pages} {013829}
		(\bibinfo {year} {2016})}\BibitemShut {NoStop}%
	\bibitem [{\citenamefont {Poladian}\ \emph {et~al.}(1991)\citenamefont
		{Poladian}, \citenamefont {Snyder},\ and\ \citenamefont
		{Mitchell}}]{Poladian91}%
	\BibitemOpen
	\bibfield  {author} {\bibinfo {author} {\bibfnamefont {L.}~\bibnamefont
			{Poladian}}, \bibinfo {author} {\bibfnamefont {A.}~\bibnamefont {Snyder}}, \
		and\ \bibinfo {author} {\bibfnamefont {D.}~\bibnamefont {Mitchell}},\ }\href
	{\doibase 10.1016/0030-4018(91)90052-F} {\bibfield  {journal} {\bibinfo
			{journal} {Opt. Commun.}\ }\textbf {\bibinfo {volume} {85}},\ \bibinfo
		{pages} {59 } (\bibinfo {year} {1991})}\BibitemShut {NoStop}%
	\bibitem [{\citenamefont {Shih}\ \emph {et~al.}(1997)\citenamefont {Shih},
		\citenamefont {Segev},\ and\ \citenamefont {Salamo}}]{Shih97}%
	\BibitemOpen
	\bibfield  {author} {\bibinfo {author} {\bibfnamefont {M.-f.}\ \bibnamefont
			{Shih}}, \bibinfo {author} {\bibfnamefont {M.}~\bibnamefont {Segev}}, \ and\
		\bibinfo {author} {\bibfnamefont {G.}~\bibnamefont {Salamo}},\ }\href
	{\doibase 10.1103/PhysRevLett.78.2551} {\bibfield  {journal} {\bibinfo
			{journal} {Phys. Rev. Lett.}\ }\textbf {\bibinfo {volume} {78}},\ \bibinfo
		{pages} {2551} (\bibinfo {year} {1997})}\BibitemShut {NoStop}%
	\bibitem [{\citenamefont {Beli\ifmmode~\acute{c}\else \'{c}\fi{}}\ \emph
		{et~al.}(1999)\citenamefont {Beli\ifmmode~\acute{c}\else \'{c}\fi{}},
		\citenamefont {Stepken},\ and\ \citenamefont {Kaiser}}]{Belic99}%
	\BibitemOpen
	\bibfield  {author} {\bibinfo {author} {\bibfnamefont {M.~R.}\ \bibnamefont
			{Beli\ifmmode~\acute{c}\else \'{c}\fi{}}}, \bibinfo {author} {\bibfnamefont
			{A.}~\bibnamefont {Stepken}}, \ and\ \bibinfo {author} {\bibfnamefont
			{F.}~\bibnamefont {Kaiser}},\ }\href {\doibase 10.1103/PhysRevLett.82.544}
	{\bibfield  {journal} {\bibinfo  {journal} {Phys. Rev. Lett.}\ }\textbf
		{\bibinfo {volume} {82}},\ \bibinfo {pages} {544} (\bibinfo {year}
		{1999})}\BibitemShut {NoStop}%
	\bibitem [{\citenamefont {Buryak}\ \emph {et~al.}(1999)\citenamefont {Buryak},
		\citenamefont {Kivshar}, \citenamefont {Shih},\ and\ \citenamefont
		{Segev}}]{Buryak99}%
	\BibitemOpen
	\bibfield  {author} {\bibinfo {author} {\bibfnamefont {A.~V.}\ \bibnamefont
			{Buryak}}, \bibinfo {author} {\bibfnamefont {Y.~S.}\ \bibnamefont {Kivshar}},
		\bibinfo {author} {\bibfnamefont {M.-f.}\ \bibnamefont {Shih}}, \ and\
		\bibinfo {author} {\bibfnamefont {M.}~\bibnamefont {Segev}},\ }\href
	{\doibase 10.1103/PhysRevLett.82.81} {\bibfield  {journal} {\bibinfo
			{journal} {Phys. Rev. Lett.}\ }\textbf {\bibinfo {volume} {82}},\ \bibinfo
		{pages} {81} (\bibinfo {year} {1999})}\BibitemShut {NoStop}%
	\bibitem [{\citenamefont {Lem-Carrillo}\ \emph {et~al.}(2014)\citenamefont
		{Lem-Carrillo}, \citenamefont {Lopez-Aguayo},\ and\ \citenamefont
		{Guti\'errez-Vega}}]{Lem-Carrillo14}%
	\BibitemOpen
	\bibfield  {author} {\bibinfo {author} {\bibfnamefont {G.}~\bibnamefont
			{Lem-Carrillo}}, \bibinfo {author} {\bibfnamefont {S.}~\bibnamefont
			{Lopez-Aguayo}}, \ and\ \bibinfo {author} {\bibfnamefont {J.~C.}\
			\bibnamefont {Guti\'errez-Vega}},\ }\href {\doibase
		10.1103/PhysRevA.90.053830} {\bibfield  {journal} {\bibinfo  {journal} {Phys.
				Rev. A}\ }\textbf {\bibinfo {volume} {90}},\ \bibinfo {pages} {053830}
		(\bibinfo {year} {2014})}\BibitemShut {NoStop}%
	\bibitem [{\citenamefont {Desyatnikov}\ \emph {et~al.}(2005)\citenamefont
		{Desyatnikov}, \citenamefont {Torner},\ and\ \citenamefont
		{Kivshar}}]{Desyatnikov05}%
	\BibitemOpen
	\bibfield  {author} {\bibinfo {author} {\bibfnamefont {A.~S.}\ \bibnamefont
			{Desyatnikov}}, \bibinfo {author} {\bibfnamefont {L.}~\bibnamefont {Torner}},
		\ and\ \bibinfo {author} {\bibfnamefont {Y.~S.}\ \bibnamefont {Kivshar}},\
	}in\ \href@noop {} {\emph {\bibinfo {booktitle} {Prog. Optics 47}}},\
	\bibinfo {editor} {edited by\ \bibinfo {editor} {\bibfnamefont
			{E.}~\bibnamefont {Wolf}}}\ (\bibinfo  {publisher} {Elsevier},\ \bibinfo
	{address} {Amsterdam},\ \bibinfo {year} {2005})\ Chap.~\bibinfo {chapter}
	{5}, pp.\ \bibinfo {pages} {291--391}\BibitemShut {NoStop}%
	\bibitem [{\citenamefont {Rotschild}\ \emph {et~al.}(2006)\citenamefont
		{Rotschild}, \citenamefont {Alfassi}, \citenamefont {Cohen},\ and\
		\citenamefont {Segev}}]{Rotschild06}%
	\BibitemOpen
	\bibfield  {author} {\bibinfo {author} {\bibfnamefont {C.}~\bibnamefont
			{Rotschild}}, \bibinfo {author} {\bibfnamefont {B.}~\bibnamefont {Alfassi}},
		\bibinfo {author} {\bibfnamefont {O.}~\bibnamefont {Cohen}}, \ and\ \bibinfo
		{author} {\bibfnamefont {M.}~\bibnamefont {Segev}},\ }\href {\doibase
		10.1038/nphys445} {\bibfield  {journal} {\bibinfo  {journal} {Nat. Phys.}\
		}\textbf {\bibinfo {volume} {2}},\ \bibinfo {pages} {1745 } (\bibinfo {year}
		{2006})}\BibitemShut {NoStop}%
	\bibitem [{\citenamefont {Solja\ifmmode \check{c}\else
			\v{c}\fi{}i\ifmmode~\acute{c}\else \'{c}\fi{}}\ \emph
		{et~al.}(1998)\citenamefont {Solja\ifmmode \check{c}\else
			\v{c}\fi{}i\ifmmode~\acute{c}\else \'{c}\fi{}}, \citenamefont {Sears},\ and\
		\citenamefont {Segev}}]{Soljacic98}%
	\BibitemOpen
	\bibfield  {author} {\bibinfo {author} {\bibfnamefont {M.}~\bibnamefont
			{Solja\ifmmode \check{c}\else \v{c}\fi{}i\ifmmode~\acute{c}\else
				\'{c}\fi{}}}, \bibinfo {author} {\bibfnamefont {S.}~\bibnamefont {Sears}}, \
		and\ \bibinfo {author} {\bibfnamefont {M.}~\bibnamefont {Segev}},\ }\href
	{\doibase 10.1103/PhysRevLett.81.4851} {\bibfield  {journal} {\bibinfo
			{journal} {Phys. Rev. Lett.}\ }\textbf {\bibinfo {volume} {81}},\ \bibinfo
		{pages} {4851} (\bibinfo {year} {1998})}\BibitemShut {NoStop}%
	\bibitem [{\citenamefont {Solja\ifmmode \check{c}\else
			\v{c}\fi{}i\ifmmode~\acute{c}\else \'{c}\fi{}}\ and\ \citenamefont
		{Segev}(2000)}]{Soljacic00}%
	\BibitemOpen
	\bibfield  {author} {\bibinfo {author} {\bibfnamefont {M.}~\bibnamefont
			{Solja\ifmmode \check{c}\else \v{c}\fi{}i\ifmmode~\acute{c}\else
				\'{c}\fi{}}}\ and\ \bibinfo {author} {\bibfnamefont {M.}~\bibnamefont
			{Segev}},\ }\href {\doibase 10.1103/PhysRevE.62.2810} {\bibfield  {journal}
		{\bibinfo  {journal} {Phys. Rev. E}\ }\textbf {\bibinfo {volume} {62}},\
		\bibinfo {pages} {2810} (\bibinfo {year} {2000})}\BibitemShut {NoStop}%
	\bibitem [{\citenamefont {Solja\ifmmode \check{c}\else
			\v{c}\fi{}i\ifmmode~\acute{c}\else \'{c}\fi{}}\ and\ \citenamefont
		{Segev}(2001)}]{Soljacic01}%
	\BibitemOpen
	\bibfield  {author} {\bibinfo {author} {\bibfnamefont {M.}~\bibnamefont
			{Solja\ifmmode \check{c}\else \v{c}\fi{}i\ifmmode~\acute{c}\else
				\'{c}\fi{}}}\ and\ \bibinfo {author} {\bibfnamefont {M.}~\bibnamefont
			{Segev}},\ }\href {\doibase 10.1103/PhysRevLett.86.420} {\bibfield  {journal}
		{\bibinfo  {journal} {Phys. Rev. Lett.}\ }\textbf {\bibinfo {volume} {86}},\
		\bibinfo {pages} {420} (\bibinfo {year} {2001})}\BibitemShut {NoStop}%
	\bibitem [{\citenamefont {Grow}\ \emph {et~al.}(2007)\citenamefont {Grow},
		\citenamefont {Ishaaya}, \citenamefont {Vuong},\ and\ \citenamefont
		{Gaeta}}]{Grow07}%
	\BibitemOpen
	\bibfield  {author} {\bibinfo {author} {\bibfnamefont {T.~D.}\ \bibnamefont
			{Grow}}, \bibinfo {author} {\bibfnamefont {A.~A.}\ \bibnamefont {Ishaaya}},
		\bibinfo {author} {\bibfnamefont {L.~T.}\ \bibnamefont {Vuong}}, \ and\
		\bibinfo {author} {\bibfnamefont {A.~L.}\ \bibnamefont {Gaeta}},\ }\href
	{\doibase 10.1103/PhysRevLett.99.133902} {\bibfield  {journal} {\bibinfo
			{journal} {Phys. Rev. Lett.}\ }\textbf {\bibinfo {volume} {99}},\ \bibinfo
		{pages} {133902} (\bibinfo {year} {2007})}\BibitemShut {NoStop}%
	\bibitem [{\citenamefont {Firth}\ and\ \citenamefont
		{Skryabin}(1997)}]{Firth97}%
	\BibitemOpen
	\bibfield  {author} {\bibinfo {author} {\bibfnamefont {W.~J.}\ \bibnamefont
			{Firth}}\ and\ \bibinfo {author} {\bibfnamefont {D.~V.}\ \bibnamefont
			{Skryabin}},\ }\href {\doibase 10.1103/PhysRevLett.79.2450} {\bibfield
		{journal} {\bibinfo  {journal} {Phys. Rev. Lett.}\ }\textbf {\bibinfo
			{volume} {79}},\ \bibinfo {pages} {2450} (\bibinfo {year}
		{1997})}\BibitemShut {NoStop}%
	\bibitem [{\citenamefont {Skryabin}\ and\ \citenamefont
		{Firth}(1998)}]{Skryabin98}%
	\BibitemOpen
	\bibfield  {author} {\bibinfo {author} {\bibfnamefont {D.~V.}\ \bibnamefont
			{Skryabin}}\ and\ \bibinfo {author} {\bibfnamefont {W.~J.}\ \bibnamefont
			{Firth}},\ }\href {\doibase 10.1103/PhysRevE.58.3916} {\bibfield  {journal}
		{\bibinfo  {journal} {Phys. Rev. E}\ }\textbf {\bibinfo {volume} {58}},\
		\bibinfo {pages} {3916} (\bibinfo {year} {1998})}\BibitemShut {NoStop}%
	\bibitem [{\citenamefont {Vinçotte}\ and\ \citenamefont
		{Bergé}(2006)}]{Vinçotte06}%
	\BibitemOpen
	\bibfield  {author} {\bibinfo {author} {\bibfnamefont {A.}~\bibnamefont
			{Vinçotte}}\ and\ \bibinfo {author} {\bibfnamefont {L.}~\bibnamefont
			{Bergé}},\ }\href {\doibase 10.1016/j.physd.2006.09.023} {\bibfield
		{journal} {\bibinfo  {journal} {Physica D}\ }\textbf {\bibinfo {volume}
			{223}},\ \bibinfo {pages} {163 } (\bibinfo {year} {2006})}\BibitemShut
	{NoStop}%
	\bibitem [{\citenamefont {Vuong}\ \emph {et~al.}(2006)\citenamefont {Vuong},
		\citenamefont {Grow}, \citenamefont {Ishaaya}, \citenamefont {Gaeta},
		\citenamefont {'t~Hooft}, \citenamefont {Eliel},\ and\ \citenamefont
		{Fibich}}]{Vuong06}%
	\BibitemOpen
	\bibfield  {author} {\bibinfo {author} {\bibfnamefont {L.~T.}\ \bibnamefont
			{Vuong}}, \bibinfo {author} {\bibfnamefont {T.~D.}\ \bibnamefont {Grow}},
		\bibinfo {author} {\bibfnamefont {A.}~\bibnamefont {Ishaaya}}, \bibinfo
		{author} {\bibfnamefont {A.~L.}\ \bibnamefont {Gaeta}}, \bibinfo {author}
		{\bibfnamefont {G.~W.}\ \bibnamefont {'t~Hooft}}, \bibinfo {author}
		{\bibfnamefont {E.~R.}\ \bibnamefont {Eliel}}, \ and\ \bibinfo {author}
		{\bibfnamefont {G.}~\bibnamefont {Fibich}},\ }\href {\doibase
		10.1103/PhysRevLett.96.133901} {\bibfield  {journal} {\bibinfo  {journal}
			{Phys. Rev. Lett.}\ }\textbf {\bibinfo {volume} {96}},\ \bibinfo {pages}
		{133901} (\bibinfo {year} {2006})}\BibitemShut {NoStop}%
	\bibitem [{\citenamefont {Silahli}\ \emph {et~al.}(2015)\citenamefont
		{Silahli}, \citenamefont {Walasik},\ and\ \citenamefont
		{Litchinitser}}]{Silahli:15}%
	\BibitemOpen
	\bibfield  {author} {\bibinfo {author} {\bibfnamefont {S.~Z.}\ \bibnamefont
			{Silahli}}, \bibinfo {author} {\bibfnamefont {W.}~\bibnamefont {Walasik}}, \
		and\ \bibinfo {author} {\bibfnamefont {N.~M.}\ \bibnamefont {Litchinitser}},\
	}\href {\doibase 10.1364/OL.40.005714} {\bibfield  {journal} {\bibinfo
			{journal} {Opt. Lett.}\ }\textbf {\bibinfo {volume} {40}},\ \bibinfo {pages}
		{5714} (\bibinfo {year} {2015})}\BibitemShut {NoStop}%
	\bibitem [{\citenamefont {Grow}\ \emph {et~al.}(2006)\citenamefont {Grow},
		\citenamefont {Ishaaya}, \citenamefont {Vuong}, \citenamefont {Gaeta},
		\citenamefont {Gavish},\ and\ \citenamefont {Fibich}}]{Grow:06}%
	\BibitemOpen
	\bibfield  {author} {\bibinfo {author} {\bibfnamefont {T.~D.}\ \bibnamefont
			{Grow}}, \bibinfo {author} {\bibfnamefont {A.~A.}\ \bibnamefont {Ishaaya}},
		\bibinfo {author} {\bibfnamefont {L.~T.}\ \bibnamefont {Vuong}}, \bibinfo
		{author} {\bibfnamefont {A.~L.}\ \bibnamefont {Gaeta}}, \bibinfo {author}
		{\bibfnamefont {N.}~\bibnamefont {Gavish}}, \ and\ \bibinfo {author}
		{\bibfnamefont {G.}~\bibnamefont {Fibich}},\ }\href {\doibase
		10.1364/OE.14.005468} {\bibfield  {journal} {\bibinfo  {journal} {Opt.
				Express}\ }\textbf {\bibinfo {volume} {14}},\ \bibinfo {pages} {5468}
		(\bibinfo {year} {2006})}\BibitemShut {NoStop}%
	\bibitem [{\citenamefont {He}\ \emph {et~al.}(2007)\citenamefont {He},
		\citenamefont {Malomed},\ and\ \citenamefont {Wang}}]{He:07}%
	\BibitemOpen
	\bibfield  {author} {\bibinfo {author} {\bibfnamefont {Y.~J.}\ \bibnamefont
			{He}}, \bibinfo {author} {\bibfnamefont {B.~A.}\ \bibnamefont {Malomed}}, \
		and\ \bibinfo {author} {\bibfnamefont {H.~Z.}\ \bibnamefont {Wang}},\ }\href
	{\doibase 10.1364/OE.15.017502} {\bibfield  {journal} {\bibinfo  {journal}
			{Opt. Express}\ }\textbf {\bibinfo {volume} {15}},\ \bibinfo {pages} {17502}
		(\bibinfo {year} {2007})}\BibitemShut {NoStop}%
	\bibitem [{\citenamefont {Ashkin}(1970)}]{Ashkin70}%
	\BibitemOpen
	\bibfield  {author} {\bibinfo {author} {\bibfnamefont {A.}~\bibnamefont
			{Ashkin}},\ }\href {\doibase 10.1103/PhysRevLett.24.156} {\bibfield
		{journal} {\bibinfo  {journal} {Phys. Rev. Lett.}\ }\textbf {\bibinfo
			{volume} {24}},\ \bibinfo {pages} {156} (\bibinfo {year} {1970})}\BibitemShut
	{NoStop}%
	\bibitem [{\citenamefont {He}\ \emph {et~al.}(1995)\citenamefont {He},
		\citenamefont {Heckenberg},\ and\ \citenamefont
		{Rubinsztein-Dunlop}}]{He1995}%
	\BibitemOpen
	\bibfield  {author} {\bibinfo {author} {\bibfnamefont {H.}~\bibnamefont
			{He}}, \bibinfo {author} {\bibfnamefont {N.~R.}\ \bibnamefont {Heckenberg}},
		\ and\ \bibinfo {author} {\bibfnamefont {H.}~\bibnamefont
			{Rubinsztein-Dunlop}},\ }\href {\doibase 10.1080/09500349514550171}
	{\bibfield  {journal} {\bibinfo  {journal} {J. Mod Opt.}\ }\textbf {\bibinfo
			{volume} {42}},\ \bibinfo {pages} {217} (\bibinfo {year} {1995})}\BibitemShut
	{NoStop}%
	\bibitem [{\citenamefont {Simpson}\ \emph {et~al.}(1996)\citenamefont
		{Simpson}, \citenamefont {Allen},\ and\ \citenamefont
		{Padgett}}]{Simpson1996}%
	\BibitemOpen
	\bibfield  {author} {\bibinfo {author} {\bibfnamefont {N.~B.}\ \bibnamefont
			{Simpson}}, \bibinfo {author} {\bibfnamefont {L.}~\bibnamefont {Allen}}, \
		and\ \bibinfo {author} {\bibfnamefont {M.~J.}\ \bibnamefont {Padgett}},\
	}\href {\doibase 10.1080/09500349608230675} {\bibfield  {journal} {\bibinfo
			{journal} {J. Mod. Opt.}\ }\textbf {\bibinfo {volume} {43}},\ \bibinfo
		{pages} {2485} (\bibinfo {year} {1996})}\BibitemShut {NoStop}%
	\bibitem [{\citenamefont {Volke-Sepulveda}\ \emph {et~al.}(2002)\citenamefont
		{Volke-Sepulveda}, \citenamefont {Garc\'es-Ch\'avez}, \citenamefont
		{Ch\'avez-Cerda}, \citenamefont {Arlt},\ and\ \citenamefont
		{Dholakia}}]{Volke-Sepulveda2002}%
	\BibitemOpen
	\bibfield  {author} {\bibinfo {author} {\bibfnamefont {K.}~\bibnamefont
			{Volke-Sepulveda}}, \bibinfo {author} {\bibfnamefont {V.}~\bibnamefont
			{Garc\'es-Ch\'avez}}, \bibinfo {author} {\bibfnamefont {S.}~\bibnamefont
			{Ch\'avez-Cerda}}, \bibinfo {author} {\bibfnamefont {J.}~\bibnamefont
			{Arlt}}, \ and\ \bibinfo {author} {\bibfnamefont {K.}~\bibnamefont
			{Dholakia}},\ }\href {http://stacks.iop.org/1464-4266/4/i=2/a=373} {\bibfield
		{journal} {\bibinfo  {journal} {J. Opt. B Quantum Semiclassical Opt.}\
		}\textbf {\bibinfo {volume} {4}},\ \bibinfo {pages} {S82} (\bibinfo {year}
		{2002})}\BibitemShut {NoStop}%
	\bibitem [{\citenamefont {Gordon}\ \emph {et~al.}(2007)\citenamefont {Gordon},
		\citenamefont {Blakely},\ and\ \citenamefont {Sinton}}]{Gordon07}%
	\BibitemOpen
	\bibfield  {author} {\bibinfo {author} {\bibfnamefont {R.}~\bibnamefont
			{Gordon}}, \bibinfo {author} {\bibfnamefont {J.~T.}\ \bibnamefont {Blakely}},
		\ and\ \bibinfo {author} {\bibfnamefont {D.}~\bibnamefont {Sinton}},\ }\href
	{\doibase 10.1103/PhysRevA.75.055801} {\bibfield  {journal} {\bibinfo
			{journal} {Phys. Rev. A}\ }\textbf {\bibinfo {volume} {75}},\ \bibinfo
		{pages} {055801} (\bibinfo {year} {2007})}\BibitemShut {NoStop}%
	\bibitem [{\citenamefont {Ashkin}\ \emph {et~al.}(1982)\citenamefont {Ashkin},
		\citenamefont {Dziedzic},\ and\ \citenamefont {Smith}}]{Ashkin:82}%
	\BibitemOpen
	\bibfield  {author} {\bibinfo {author} {\bibfnamefont {A.}~\bibnamefont
			{Ashkin}}, \bibinfo {author} {\bibfnamefont {J.~M.}\ \bibnamefont
			{Dziedzic}}, \ and\ \bibinfo {author} {\bibfnamefont {P.~W.}\ \bibnamefont
			{Smith}},\ }\href {\doibase 10.1364/OL.7.000276} {\bibfield  {journal}
		{\bibinfo  {journal} {Opt. Lett.}\ }\textbf {\bibinfo {volume} {7}},\
		\bibinfo {pages} {276} (\bibinfo {year} {1982})}\BibitemShut {NoStop}%
	\bibitem [{\citenamefont {Smith}\ \emph {et~al.}(1982)\citenamefont {Smith},
		\citenamefont {Maloney},\ and\ \citenamefont {Ashkin}}]{Smith:82}%
	\BibitemOpen
	\bibfield  {author} {\bibinfo {author} {\bibfnamefont {P.~W.}\ \bibnamefont
			{Smith}}, \bibinfo {author} {\bibfnamefont {P.~J.}\ \bibnamefont {Maloney}},
		\ and\ \bibinfo {author} {\bibfnamefont {A.}~\bibnamefont {Ashkin}},\ }\href
	{\doibase 10.1364/OL.7.000347} {\bibfield  {journal} {\bibinfo  {journal}
			{Opt. Lett.}\ }\textbf {\bibinfo {volume} {7}},\ \bibinfo {pages} {347}
		(\bibinfo {year} {1982})}\BibitemShut {NoStop}%
	\bibitem [{\citenamefont {El-Ganainy}\ \emph
		{et~al.}(2007{\natexlab{a}})\citenamefont {El-Ganainy}, \citenamefont
		{Christodoulides}, \citenamefont {Rotschild},\ and\ \citenamefont
		{Segev}}]{El-Ganainy:07b}%
	\BibitemOpen
	\bibfield  {author} {\bibinfo {author} {\bibfnamefont {R.}~\bibnamefont
			{El-Ganainy}}, \bibinfo {author} {\bibfnamefont {D.~N.}\ \bibnamefont
			{Christodoulides}}, \bibinfo {author} {\bibfnamefont {C.}~\bibnamefont
			{Rotschild}}, \ and\ \bibinfo {author} {\bibfnamefont {M.}~\bibnamefont
			{Segev}},\ }\href {\doibase 10.1364/OE.15.010207} {\bibfield  {journal}
		{\bibinfo  {journal} {Opt. Express}\ }\textbf {\bibinfo {volume} {15}},\
		\bibinfo {pages} {10207} (\bibinfo {year} {2007}{\natexlab{a}})}\BibitemShut
	{NoStop}%
	\bibitem [{\citenamefont {Lee}\ \emph {et~al.}(2009)\citenamefont {Lee},
		\citenamefont {El-Ganainy}, \citenamefont {Christodoulides}, \citenamefont
		{Dholakia},\ and\ \citenamefont {Wright}}]{Lee:09}%
	\BibitemOpen
	\bibfield  {author} {\bibinfo {author} {\bibfnamefont {W.~M.}\ \bibnamefont
			{Lee}}, \bibinfo {author} {\bibfnamefont {R.}~\bibnamefont {El-Ganainy}},
		\bibinfo {author} {\bibfnamefont {D.~N.}\ \bibnamefont {Christodoulides}},
		\bibinfo {author} {\bibfnamefont {K.}~\bibnamefont {Dholakia}}, \ and\
		\bibinfo {author} {\bibfnamefont {E.~M.}\ \bibnamefont {Wright}},\ }\href
	{\doibase 10.1364/OE.17.010277} {\bibfield  {journal} {\bibinfo  {journal}
			{Opt. Express}\ }\textbf {\bibinfo {volume} {17}},\ \bibinfo {pages} {10277}
		(\bibinfo {year} {2009})}\BibitemShut {NoStop}%
	\bibitem [{\citenamefont {El-Ganainy}\ \emph
		{et~al.}(2007{\natexlab{b}})\citenamefont {El-Ganainy}, \citenamefont
		{Christodoulides}, \citenamefont {Musslimani}, \citenamefont {Rotschild},\
		and\ \citenamefont {Segev}}]{El-Ganainy:07}%
	\BibitemOpen
	\bibfield  {author} {\bibinfo {author} {\bibfnamefont {R.}~\bibnamefont
			{El-Ganainy}}, \bibinfo {author} {\bibfnamefont {D.~N.}\ \bibnamefont
			{Christodoulides}}, \bibinfo {author} {\bibfnamefont {Z.~H.}\ \bibnamefont
			{Musslimani}}, \bibinfo {author} {\bibfnamefont {C.}~\bibnamefont
			{Rotschild}}, \ and\ \bibinfo {author} {\bibfnamefont {M.}~\bibnamefont
			{Segev}},\ }\href {\doibase 10.1364/OL.32.003185} {\bibfield  {journal}
		{\bibinfo  {journal} {Opt. Lett.}\ }\textbf {\bibinfo {volume} {32}},\
		\bibinfo {pages} {3185} (\bibinfo {year} {2007}{\natexlab{b}})}\BibitemShut
	{NoStop}%
	\bibitem [{\citenamefont {Man}\ \emph {et~al.}(2013)\citenamefont {Man},
		\citenamefont {Fardad}, \citenamefont {Zhang}, \citenamefont {Prakash},
		\citenamefont {Lau}, \citenamefont {Zhang}, \citenamefont {Heinrich},
		\citenamefont {Christodoulides},\ and\ \citenamefont {Chen}}]{Man13}%
	\BibitemOpen
	\bibfield  {author} {\bibinfo {author} {\bibfnamefont {W.}~\bibnamefont
			{Man}}, \bibinfo {author} {\bibfnamefont {S.}~\bibnamefont {Fardad}},
		\bibinfo {author} {\bibfnamefont {Z.}~\bibnamefont {Zhang}}, \bibinfo
		{author} {\bibfnamefont {J.}~\bibnamefont {Prakash}}, \bibinfo {author}
		{\bibfnamefont {M.}~\bibnamefont {Lau}}, \bibinfo {author} {\bibfnamefont
			{P.}~\bibnamefont {Zhang}}, \bibinfo {author} {\bibfnamefont
			{M.}~\bibnamefont {Heinrich}}, \bibinfo {author} {\bibfnamefont {D.~N.}\
			\bibnamefont {Christodoulides}}, \ and\ \bibinfo {author} {\bibfnamefont
			{Z.}~\bibnamefont {Chen}},\ }\href {\doibase 10.1103/PhysRevLett.111.218302}
	{\bibfield  {journal} {\bibinfo  {journal} {Phys. Rev. Lett.}\ }\textbf
		{\bibinfo {volume} {111}},\ \bibinfo {pages} {218302} (\bibinfo {year}
		{2013})}\BibitemShut {NoStop}%
	\bibitem [{\citenamefont {Fardad}\ \emph {et~al.}(2013)\citenamefont {Fardad},
		\citenamefont {Mills}, \citenamefont {Zhang}, \citenamefont {Man},
		\citenamefont {Chen},\ and\ \citenamefont {Christodoulides}}]{Fardad:13}%
	\BibitemOpen
	\bibfield  {author} {\bibinfo {author} {\bibfnamefont {S.}~\bibnamefont
			{Fardad}}, \bibinfo {author} {\bibfnamefont {M.~S.}\ \bibnamefont {Mills}},
		\bibinfo {author} {\bibfnamefont {P.}~\bibnamefont {Zhang}}, \bibinfo
		{author} {\bibfnamefont {W.}~\bibnamefont {Man}}, \bibinfo {author}
		{\bibfnamefont {Z.}~\bibnamefont {Chen}}, \ and\ \bibinfo {author}
		{\bibfnamefont {D.~N.}\ \bibnamefont {Christodoulides}},\ }\href {\doibase
		10.1364/OL.38.003585} {\bibfield  {journal} {\bibinfo  {journal} {Opt.
				Lett.}\ }\textbf {\bibinfo {volume} {38}},\ \bibinfo {pages} {3585} (\bibinfo
		{year} {2013})}\BibitemShut {NoStop}%
	\bibitem [{\citenamefont {Steblina}\ \emph {et~al.}(1998)\citenamefont
		{Steblina}, \citenamefont {Kivshar},\ and\ \citenamefont
		{Buryak}}]{Steblina:98}%
	\BibitemOpen
	\bibfield  {author} {\bibinfo {author} {\bibfnamefont {V.~V.}\ \bibnamefont
			{Steblina}}, \bibinfo {author} {\bibfnamefont {Y.~S.}\ \bibnamefont
			{Kivshar}}, \ and\ \bibinfo {author} {\bibfnamefont {A.~V.}\ \bibnamefont
			{Buryak}},\ }\href {\doibase 10.1364/OL.23.000156} {\bibfield  {journal}
		{\bibinfo  {journal} {Opt. Lett.}\ }\textbf {\bibinfo {volume} {23}},\
		\bibinfo {pages} {156} (\bibinfo {year} {1998})}\BibitemShut {NoStop}%
\end{thebibliography}
\end{document}